# Break-down of the relationship between $\alpha$-relaxation and equilibration in hydrostatically compressed metallic glasses

**Antoine Cornet[1,3*†] Jie Shen[1,3*†], Alberto Ronca[1,3], Shubin Li[2,3], Nico Neuber[4], Maximilian Frey[4], Eloi Pineda[5], Thierry Deschamps[2], Christine Martinet[2], Sylvie Le Floch[2], Daniele Cangialosi[6,7], Yuriy Chushkin[3], Federico Zontone[3], Marco Cammarata[3], Gavin B. M. Vaughan[3], Marco di Michiel[3], Gaston Garbarino[3], Ralf Busch[4], Isabella Gallino[8], Celine Goujon[1], Murielle Legendre[1], Geeth Manthilake[9], and Beatrice Ruta[1,3*]**

[1] *Institut Néel, Université Grenoble Alpes and Centre National de la Recherche Scientifique, 25 rue des Martyrs - BP 166, 38042, Grenoble cedex 9 France*

[2] *Institut Lumière Matière UMR CNRS 5306, Université Claude Bernard Lyon 1, CNRS, F-69622 Villeurbanne, France*

[3] *European Synchrotron Radiation Facility, 71 avenue des Martyrs, CS 40220, Grenoble 38043, France*

[4] *Chair of Metallic Materials, Saarland University, Campus C6.3, 66123 Saarbrücken, Germany*

[5] *Department of physics, Institute of Energy Technologies, Universitat Politècnica de Catalunya-BarcelonaTech, 08019 Barcelona, Spain*

[6] *Donostia International Physics Center, Paseo Manuel de Lardizabal 4, 20018 San Sebastián, Spain*

[7] *Centro de Física de Materiales (CSIC–UPV/EHU), Paseo Manuel de Lardizabal 5, 20018 San Sebastián, Spain*

[8] *Chair of Metallic Materials, Berlin Institute of Technology (TU-Berlin), Ernst-Reuter Platz 1, 10587 Berlin, Germany*

[9] *Laboratoire Magmas et Volcans CNRS, IRD, OPGC, Université Clermont Auvergne, 53000 Clermont-Ferrand, France*

[†] These authors equally contributed to the work

*Corresponding authors

antoine.cornet@neel.cnrs.fr

jie.shen@neel.cnrs.fr

beatrice.ruta@neel.cnrs.fr

## Abstract

Glasses encode the memory of any thermo-mechanical treatment applied to them. This ability is associated to the existence of a myriad of metastable amorphous states which can be probed through different experimental pathways. It is usually assumed that this memory can be erased in the supercooled liquid, and that this process occurs on a time scale controlled by the $\alpha$-relaxation. We find that this assumption does not apply for hydrostatically compressed glasses. Annealing under pressure a prototypical metallic glass can irreversibly modify its dynamics, thermodynamics and structure, reduce the atomic mobility and lead to structural modifications of the first coordination shells which reduce the thermal stability with respect to a glass annealed in absence of pressure. When heated above their glass transition temperature, these compressed glasses do not convert into the pristine supercooled liquid, implying the existence of an additional process, beyond the $\alpha$-relaxation, contributing to the equilibrium recovery of the material. These results establish pressure as a powerful tool for engineering non-equilibrium glassy materials with tailored properties, while deepening our understanding of relaxation dynamics in disordered systems under extreme conditions.

## Introduction

A glass is a mechanically stable solid which is usually obtained by cooling a liquid fast enough to avoid crystallization. The resulting material has inherited the disorder and isotropic structure of the corresponding liquid and keeps the memory of any thermal and mechanical protocol applied to it. As a large variety of materials can become glasses, the glass state has been considered as a possible fourth state of matter[1,2].

The possibility to drive a glass through different metastable states by applying thermal and mechanical treatments has fascinating scientists for decades as it is also crucial for the design of novel systems with extended service life[3,4]. Despite the countless works in the field, a microscopic understanding of the underlying structural and dynamical mechanisms involved in amorphous driven materials remains to this day an open problem.

Hydrostatic compression has recently emerged as a powerful tool to tune the properties of metallic glasses (MGs) with promising potential applications[5–7]. While compression leads to relaxation in oxide glasses[8,9] and thus to a lower energetic configuration, studies in MGs have reported contrasting results depending on the material and the applied compression protocol. Contrary to the intuitive expectation that high-pressure annealing should produce a denser and more relaxed glassy state, some systems (as for instance La-based MGs) undergo thermal rejuvenation[7], as indicated by an increase in fictive temperature ($T_f$), corresponding to a transition to a higher-energy state in the potential energy landscape (PEL)[5]. A surprising volume expansion and thermal rejuvenation has been reported also after tri-axial compression in Zr-based MGs at ambient temperature, and appears related to a non-monotonous pressure-induced acceleration of the atomic motion[10], likely associated to the emergence of a secondary local minima in the PEL[11].

In contrast, high-pressure annealing of other Zr- and Pd-based MGs has been found to produce relaxation and enhanced stability[12,13] as in other families of glass formers[14–16]. Our recent study shows that the pressure response of $Zr_{46.8}Ti_{8.2}Cu_{7.5}Ni_{10}Be_{27.5}$ alloys depends on the microscopic relaxation mechanism active during the compression which determines the structural and energetic states of MGs[17].

The above discussed studies focus however only on the properties of recovered compressed MGs at ambient conditions and almost nothing is known on the thermal response of these materials and on the microscopic mechanisms accompanying their devitrification in the supercooled liquid phase in the absence of mechanical constraints. There is a general consensus that the memory of the previous thermal and mechanical protocols applied to a glass can be erased by heating the material in the supercooled liquid phase and that this recovery of the original state is controlled by the main α-structural relaxation process[18,19]. Following this assumption, as-cast glasses are often heated close to their glass transition temperature, $T_g$, to erase their history prior to use them in applications or for further characterizations. While this approach has been extensively tested and verified in thermally treated glasses[18–20], the microscopic mechanisms accompanying the release of the constraints imposed by high pressure compressions and the consequent equilibrium recovery upon heating under ambient pressure have not been investigated so far.

In order to rationalize previous results and get more insights into the effect played by hydrostatic compression in MGs and their equilibrium restoring during devitrification, we present here a detailed experimental investigation of the structure, dynamics and thermodynamic state of a series of $Pt_{42.5}Cu_{27}Ni_{19.5}P_{21}$ glasses obtained by compressing the pristine material in the GPa pressure range and at temperatures across its glass transition. Contrary to the intuitive idea of a more compacted and relaxed state, we show that annealing a glass at high pressure lowers its thermal stability with respect to a treatment in absence of pressure, even in the proximity of the glass transition. This effect is accompanied by a surprising structural expansion of the first coordination shells which persists up to length scales of ~9Å, resulting in a combination of thermal expansion and contraction at different length scales. When heated above their glass transition temperature in the absence of any mechanical constraints, the structure and dynamics of these compressed glasses do not convert into the pristine system, challenging the generally accepted idea that the memory of any thermo-mechanical treatment applied to a glass can be rapidly erased in the supercooled liquid on a time scale dictated by the structural α-relaxation process. These results imply a complex devitrification mechanism after compression and a non-exclusive relationship between equilibration and α-relaxation.

## Results

**Influence of the physical state reached by the system during compression.** In order to identify the experimental conditions necessary to lower the thermal stability of MGs by means of hydrostatic compression, we performed different densifications protocols varying the compression temperature $T_{comp}$ through the pressure dependent glass transition temperature $T_g(P)$ at a constant pressure of 5 GPa (Fig. 1a). A summary of all samples preparations is reported in the Methods and SM. In the following we will call *high pressure quenched glasses* (HPQGs) those glasses obtained by cooling from the supercooled liquid under pressure at $T_{comp} > T_g(P)$, and *high-pressure annealed* *glasses* (HPAGs) those compressed by heating the glass at high pressure at $T_{\mathrm{comp}} < T_{\mathrm{g}}(P)$. Although the exact value of $T_{\mathrm{g}}(P)$ is not known *at priori*, following previous works[13,21], we estimate an increase of $T_{\mathrm{g}}$ of ~20 K in 5 GPa and that heating to $T_{\mathrm{comp}}$~$T_{\mathrm{g}}$(1atm)+50 K should assure to reach the supercooled liquid phase under compression, as confirmed also by our thermal analysis discussed below (see also Methods). To avoid crystallization, pressure is first applied to the glass, then the temperature is raised under isobaric conditions to the target value. With our procedure, which is described in detail in Methods, the resulting glass retains information about the inelastic deformation of the high pressure and temperature state. The thermodynamic response of all compressed glasses was then measured using a novel approach consisting of single-shot flash differential scanning calorimetry (FDSC), i.e. without the usual melting of the material on the chip[22] and the results are summarized in Table 1.

The kinetic response of glasses compressed at 5GPa from different thermodynamic states is shown in Fig. 1b together with data of the reference system subjected to the same thermal treatment but at 1 atm. The largest recovery is shown by the HPQG that was prepared by cooling the corresponding highly dense liquid. This means the HPQG is more relaxed and has enhanced kinetic stability with respect to the reference, and implies a slower dynamics of the corresponding HP liquid, in agreement with numerical simulations[23] and experimental studies in other glass-formers[14]. This relaxation is represented by the highest enthalpy peak and an increased $T_{\mathrm{g,onset}}$ of about 12 K. This value is consistent with those reported in literature for other MGs, which were estimated using enthalpy recovery methods[12,13]. In stark contrast, annealing the glass at high pressure leads to a less stable state with respect to the reference sample annealed at 1 atm, even in the proximity of $T_{\mathrm{g}}$. As shown in Fig. 1b, the kinetic response of a HPAG at 5GPa obtained from compression at $T_{\mathrm{comp}}$=0.98$T_{\mathrm{g}}$(5 GPa) has lower enthalpy recovery and a negative shift of $T_{\mathrm{g,onset}}$ of about 9 K. This value is furthermore independent of the applied heating rate in the 100-1000 K/s range (Fig. S1). The enthalpy recovery peak of the HPAG is a simple consequence of the fast heating and its lower value with respect to the reference strengthens the occurrence of a thermodynamically less stable state. This profile corresponds to a standard exothermic contribution below $T_g$ when measured with the same heating and cooling rate (Fig. S2). The different stability of the compressed glasses is confirmed by their fictive temperatures, $T_f$, reported in Table 1 that we evaluated using the equal-area construction[24] (Fig. 1c, 1d and S3).

**Structural characterization of HPAG.** While enhanced thermal stability is common in high-pressure quenched glasses[14], the almost ubiquitous lower stability of high-pressure annealed MGs with respect to simple annealing treatments is less intuitive[7,25]. This behavior is often accompanied by structural rejuvenation, i.e. a pressure induced expansion of the structure with respect to the pristine material[5]. To better understand the lower stability in HPAGs and the physical mechanisms occurring during the equilibrium recovery, we performed several synchrotron X-ray Diffraction (XRD) experiments to probe the structure not only after the compression, but also during heating in the liquid and subsequent cooling at ambient pressure, thus without applying any mechanical constraint. The behavior of the first sharp diffraction peak (FSDP) in the static structure factor, $S(q)$, is shown in Fig. 2a, Fig. S4 and Fig. S5. To highlight the structural changes induced by the high-pressure annealing, the data are compared to the extreme case of a hyper-quenched glass (i.e. obtained by a fast-cooling rate of $10^5$ K/s from the melt, see Methods section). The latter can be viewed as an example of a structurally rejuvenated glass, as it exhibits extra free volume with respect to the reference state obtained after thermal cycling across $T_g$ at low rates.

A non-monotonic temperature dependence of the FSDP in the HPAG is evident by simple visual inspection and is quantified in Fig. 2b and 2c by the evolution of the inverse position of the FSDP, $1/q_{FSDP}^3$ and of the full width half maximum (FWHM). After compression, the HPAG exhibits structural expansion with respect to the hyper-quenched glass, as signaled by the larger value of $1/q_{FSDP}^3$, a quantity often associated with the volume as being representative of the medium range order (MRO)[26]. At the same time the lower value of the FWHM implies a more ordered structure. On heating, the structure expands until the glass transition region, where significant thermally activated structural reorganization occurs. Here, the major constraints imposed by the densification protocol are released with a surprising volume contraction, before the usual thermal expansion of the liquid above $T_g$ (~520 K). This behavior mirrors that of the hyper-quenched glass, where the extra free volume associated to the fast-quenching is released when the glass is heated in the proximity of $T_g$. The two systems however differ at the level of the FWHM. While the compressed glass becomes more disordered close to the glass transition, the release of extra free volume in the hyper-quenched glass is accompanied by structural ordering. The same conclusion is obtained by independent measurements of the evolution of the peak intensity of neighboring shells in the pair distribution function (PDF) in Fig. 3a and Fig. 3b. For $r$>5 Å, the peak intensity in the hyper-quenched glass increases with temperature due to increased ordering resulting from relaxation on heating. In the compressed glass, on the other hand, the peak intensity decreases suggesting a transformation into a more packed but disordered structure during devitrification (see also Fig. S6 and S7).

The structural mechanisms occurring in both samples can be discussed at the light of the known structure of $Pt_{42.5}Cu_{27}Ni_{9.5}P_{21}$ metallic glass, in particular in the different relevant clusters of atoms and their connectivity. Such clusters include the 5 fold symmetry clusters such as icosahedron typical of metallic melts[27] which develop further upon undercooling[28,29], as well as more complex clusters which are due to the presence of semi-metals in metal-metalloid alloys, where the complex and partially covalent bonding leads to directional bonds and new types of clusters[30]. To the best of our knowledge, there are no numerical simulations identifying the relevant (partially covalent) topological clusters representative of the $Pt_{42.5}Cu_{27}Ni_{9.5}P_{21}$ glass structure, but extensive literature exists on the closely related PdCuNiP alloys and the PdNiP and PdCuP subsystems[31–36]. These reports provide a clear picture with two P-centered types of clusters: caped trigonal prisms and tetragonal dodecahedral clusters, the former involving mainly Ni atoms and the latter mainly Cu atoms[31,32]. The coexistence of both types of clusters (metallic based icosahedron and partially covalent prisms and dodecahedron) has been suggested from ab-initio simulations supporting experimental structural studies also in PdNiP[36]. Finally, the link between the structure of the $Pd_{42.5}Cu_{27}Ni_{9.5}P_{21}$ and $Pt_{42.5}Cu_{27}Ni_{9.5}P_{21}$ liquids and glasses was investigated by systematic substitution of Pd atoms by Pt atoms and subsequent structural characterization by XRD[37]. At the cluster level, this showed a limited icosahedral proportion in $Pt_{42.5}Cu_{27}Ni_{9.5}P_{21}$ with respect to $Pd_{42.5}Cu_{27}Ni_{9.5}P_{21}$.

The connectivity between these clusters is encoded in the second coordination shell[38]. Depending on the number of atoms shared by two adjacent clusters, the most probable distances for all connection schemes can be expressed as a function of the nearest-neighbor distance $r_1$: $2r_1$, $\sqrt{3}r_1$, $\sqrt{8/3}\,r_1$ and $\sqrt{2}r_1$ for 1 atom (vertex), 2 atoms (edge), 3 and 4 atoms (face) connectivity respectively. As shown in Fig. 4a, the differential pair distribution function shows relative intensity changes correlated with these most probable distances, indicating different changes in the cluster connectivity. For the hyper-quenched glass, the structural ordering on heating toward $T_g$ primarily originates from free volume release associated with structural relaxation. In the $Pt_{42.5}Cu_{27}Ni_{9.5}P_{21}$ system, this ordering is reflected in the further increase of the 3-atoms connection mode in the as-cast sample, indicating a packed structure which can be associate to an increased connectivity of icosahedral clusters[37,39]. Meanwhile, we observe also a slight increase in the 2-atoms connection mode which has been associated to P-centered clusters (trigonal-prisms and dodecahedron) in previous works by Gross *et al*[37].

Even if a direct connection between the different scheme modes and the densified structure is more complex to carry on due to the lack of simulations and complementary works under pressure, we can use atmospheric pressure analysis to associate the different structural motifs to the observed changes in the second coordination shell. In contrast to the hyper-quenched glass, in the HPAG sample, both 2-atoms and 3-atoms connection modes decrease with increasing temperature, suggesting a less packed structure on releasing the pressure with a possible reduction in icosahedral and trigonal-prisms/dodecahedron clusters[37]. At the same time, there is continuous increase in the 4-atom connection mode which implies a local structural reorganization that cannot be directly associated to the two main structural motifs based on the current state of the literature in the field.

We note that the 3-atoms and 4-atoms connections exhibit opposite temperature evolutions in the two systems, mirroring the opposite trend of the FWHM of the FSDP and of the intensity of the PDF peaks (Fig. 2 and S6). This suggests that the increasing disorder in the HPAG on heating could arise from the decreasing of icosahedral connections and the increasing in the 4-atom connection mode, which suggests the formation of octahedrons on heating. Additional experiments and simulations will be necessary to identify more accurately the corresponding structural motifs.

The PDF analysis shows also that the structural changes are accompanied by an expanded structure with respect to the hyper-quenched glass up to ~9 Å only (Fig. 4b). At larger distances, the HPAG structure is more compact, as one would intuitively expect after densification (Fig. S8). This effect results in the surprising concomitance of negative and positive thermal expansions of the MRO, as indicated by the temperature dependence of the distance of neighboring shells $r_3$, $r_4$ and $r_5$ upon heating close to $T_g$.

**Absence of recovery of the original state.** It is a common assumption that any previous thermo-mechanical treatment applied to a glass can be erased by heating the material into the supercooled liquid. Although above $T_g$ both systems exhibit similar thermal expansions (Fig. 4a), the HPAG transforms into a more structurally compact state, with shorter distances at all neighboring shells, which persist even after subsequent cooling in the glass. This counterintuitive result suggests the possibility of obtaining new states of the same composition after releasing the constraints induced by the high-pressure annealing.

The structural dissimilarities between these two states can again be discussed with respect to the known structures of the glass and its experimental signatures, including the cluster connectivity and in particular the cross-samples differential pair distribution function $G_{ref}(r,T) - G_{HPA}(r,T)$ (Fig. 5), which shows that the structural differences in the connection schemes between the hyper-quenched and HPA samples decrease on heating, but do not vanish. This is in line with the persisting difference in density revealed by the mean correlation distances of the successive coordination shells (Fig. 4a). In Fig. 5 negative values correspond to dominant features in the HPAG. While there are almost no differences in the 1-atom connections in the two samples above $T_g$, differences in the connection schemes remain for the 2 to 4- atoms connections. In particular in the HPA sample there are more 2- and 3-atoms connections which, using again atmospheric pressure analysis data[37], could be associated to a large number of P-centered clusters (trigonal prisms and dodecahedron) and icosahedron, with the largest difference in the P-centered clusters. Differently, the 4-atoms connection scheme, although increasing during the pressure release in the HPAG, remains lower than the reference liquid, suggesting a lower restoring of octahedrons once the constrains imposed by the compression are removed

This unexpected irreversibility is confirmed by dynamical data, probed by X-Ray Photon Correlation Spectroscopy (XPCS). This technique allows to get information on the evolution of the intermediate scattering function (ISF) $F(q,t)$ through the computation of the temporal intensity-intensity correlation function $g^{(2)}(q,t)$, where $q$ refers to the wave-vector and $t$ to time, via the Siegert relation $g^{(2)}(q,t) = 1 + \gamma|F(q,t)|^2$, where $\gamma$ is the experimental contrast[40]. Figure 6 shows the dynamics of the supercooled liquids obtained upon heating HPAG samples previously annealed at 5 and 10 GPa together with reference data at 1 atm. A typical correlation function spans up to 4 decades in time (Fig. 6a), indicating a stretched exponential decay which can be modelled using the Kohlrausch-Williams-Watts (KWW) function: $|F(q,t)|^2 = exp\left[-2(t/\tau_\alpha\,(T,P))^{\beta(T,P)}\right]$, with $\tau_\alpha$ the structural relaxation time and $\beta(T,P)<1$ the shape exponent describing the degree of heterogeneity of the dynamics.

When heated above $T_g$ at ambient pressure, the HPAG transforms into a new system with respect to the reference, with faster decay times in the ISFs (Fig. 6a) which corresponds to lower relaxation times (Fig. 6b), implying a variation $\left(\frac{dTg}{dP}\right)_{ex-situ} < 0$, in agreement with the FDSC results. This faster dynamic can be further accelerated by increasing the pressure during the densification protocol, as shown by data on a second HPAG produced with the same protocol but at 10 GPa (Fig. 6a and 6b), which exhibits also lower thermal stability (see Fig. S9). The different dynamics in the three systems confirm the absence of recovery of the original supercooled liquid in agreement with the structural measurements. In addition, it is worth noting that our data reveal that the magnitude of the dynamic differences between the different states varies with the compression pressure, and it remains unclear whether a critical pressure below 5 GPa exists at which these dynamic differences begin to emerge. As discussed in the Methods, probing larger pressure ranges is technically challenging in MGs and require further experiments and higher XPCS resolution.

We eliminate the possibility of a continuous reverse transition towards the original liquid, as all data corresponds to stationary dynamics over more than ≈ 30,000 s which is much than $\tau_\alpha$. As each data point is measured during an isotherm, we can check the perfectly stationary dynamics from the Two-Times Correlation Functions (TTCFs) measured using XPCS (Fig. 6c). These functions are a time-resolved evolution of the ISF. The width of the profile along the diagonal is proportional to $\tau_\alpha$ and its constant value indicates steady dynamics, evidence of equilibrium at the probed length scale.

**Discussion**

Our work shows that the properties of *ex-situ* compressed glasses of $Pt_{42.5}Cu_{27}Ni_{19.5}P_{21}$ can be tuned by varying the nature of the state attained during the compression. While densification from a supercooled liquid leads to a more relaxed glass, the stability of the material can be dramatically lowered by compressing in the glass, even in the proximity of $T_g$. This contrasting pressure response probably arises from the different mechanisms of atomic motion in the two states[41–43]. In the supercooled liquid, the dynamics is heterogeneous and associated with viscous flow[44,45]. The configurationally frozen glassy state instead is governed by super-diffusive[10] particle displacement which originates from long-range elastic interactions[46–48] and heterogeneous cluster dynamics[49]. These results are also confirmed by our recent work where we show that when β-relaxation is the dominating process (associated with localized atomic motions), structural disorder increases without substantial density changes leading to an increase in $T_f$. Differently, when α-relaxation dominates (involving cooperative atomic rearrangements), densification and structural ordering are promoted, thereby increasing the stability and lowering the $T_f$[17].

Although HPQGs have been widely investigated and are generally associated with enhanced thermal stability due to the formation of a densely packed structure, our study deliberately focuses on HPAGs, which exhibit far less intuitive behaviors. Unlike HPQGs, HPAGs are produced by applying pressure to the glassy state at temperatures below $T_{g,P}$, where atomic mobility is strongly restricted due to the reduced free volume under pressure and the thermal energy not large enough to activate the structural relaxation, with a delay of relaxation to higher temperatures. This unique sample preparation condition makes HPAGs an ideal system for probing the limits of glass structure and thermodynamic states under high pressure. Surprisingly, instead of becoming more relaxed as one might expect, HPAGs exhibit higher $T_f$ and a certain degree of atomic-scale structural expansion with respect to samples annealed at ambient pressure. It must be noted that this process is not similar to pure thermal rejuvenation of the glass, as its structure is not compatible with an evolution towards a more unrelaxed state (see Fig. 2 and 3).

The structural changes in HPAG are reflected in the increased structural order and an unexpected expansion of the SRO and MRO up to about 9 Å, which is followed by densification at longer distances. To our surprise, these compressed glasses do not convert in the original liquid when heated above $T_g$ at ambient pressure. Although the structural modifications induced by the densification protocol can be largely relaxed on approaching $T_g$, HPAGs transform into new systems with respect to the original liquid with faster relaxation times, a more disordered and packed structure, characterized by a larger number of P-centered clusters (trigonal prisms and dodecahedron) and icosahedrons. This irreversibility challenges the common belief that the memory of any thermo-mechanical treatment applied to a glass can be erased above $T_g$.

A possible interpretation is that a polyamorphic transition occurs during the compression protocol and that the HPAGs convert into new supercooled liquids when heated above their $T_g$. During hot compression in the glassy state, the structure is modified which could suggest the occurrence of a pressure-induced glass–glass transition, generating a new glass. This would be consistent with the structural differences between the two systems shown in Fig. 3. Such a new glass, "if heated while still under pressure", could in principle form a new liquid once in equilibrium, but this is not done in our work. In the present case, the HPAGs are heated at ambient pressure "without any additional constraint". If we assume that our data represent different supercooled liquids of the same composition, this interpretation would imply the possibility of obtaining multiple liquids at the same (P,T) point in the phase diagram in thermodynamic equilibrium, as our data are measured at ambient pressure and in the same temperature range. This scenario would imply the co-existence of at least three liquids (Fig. 6), or even more by varying the compression pressure, $P_{comp}$. To the best of our knowledge, this interpretation looks unphysical and only the co-existence of two different liquids has been observed so far in the presence of some liquid-liquid transitions (LLTs)[50,51]. So, we exclude LLTs as a possible interpretation of the lack of recovery above $T_g$. The only possible reason of the persistence of a memory of the previous treatments is that the materials are still out of equilibrium even if heated above their glass transition temperature. At the same time, the fact that we observe liquid features such as a positive thermal expansion of the MRO and stable supercooled liquid dynamics at atomic scales, suggests that the system is in equilibrium when regarded over spatial distances covering few tens of Å, as those probed by XPCS and XRD/PDF. The only possible explanation to reconcile all these arguments is the absence of a full equilibrium recovery and the presence of a length scale dependent devitrification process. This would be the consequence of the fact that both XPCS and XRD probe microscopic scales and cannot ensure that the material is macroscopically in equilibrium. In this scenario, the system is still out of equilibrium and is trapped in a different basin of the potential energy landscape with respect to the equilibrium liquid (which is the reference liquid at 1 atm).

This result contradicts the paradigmatic view that the $\alpha$-relaxation controls the devitrification of a glass into a supercooled liquid, inducing a rapid recovery of standard behavior on the same time scale as $\tau_\alpha$, as it is usually the case when only thermal treatments are applied to a glass[41]. This is not observed in our experiments, as the atomic dynamics remains stationary over experimental time scales larger than $10^3$ $\tau_\alpha$, without any signature of recovery. This breakdown of the relationship between $\alpha$-relaxation and equilibration in the supercooled liquid may be attributed to the permanent destruction of some important atomic clusters under high pressure (corresponding to the 4-atom connection mode) as well as an increase in icosahedral and P-centered clusters (trigonal prisms and dodecahedron) clusters. This is reflected in the contrasting responses of different neighboring shells to pressure and the opposite trend in atomic structural evolution compared to the hyper-quenched sample during the equilibrium recovery (Fig. 4a and Fig. 5). In this context, it is then more reasonable to think that the microscopic α-relaxation is not the only process involved in the equilibration and that an additional mechanism should exist, slower than $\tau_\alpha$, which contributes to the equilibration.

Interestingly, we find some similarities between HPAGs and the temperature evolution of high-density amorphous ice (HDA). HDA also converts into a liquid with faster dynamics and lower $T_g$ than the low-density amorphous (LDA) phase[50,52], and the original state is not recovered under temperature cycling across $T_g$[53]. In stark contrast, polyamorphism in water is accompanied by important structural changes and thermodynamic anomalies[54] while in our sample, pressure induces only subtle structural changes. Additional studies in MGs showing polyamorphic[55,56] transitions could help clarifying these similarities. Furthermore, our results suggest that the liquid nature of an amorphous compound might not be so trivial to establish, and that a microscopic approach is not always enough to demonstrate the thermodynamic equilibrium state of the system, at least in metallic systems in the proximity of the glass transition region after being subjected to pressure and temperature treatments.

## Materials and Methods

**Material synthesis:** We prepared a PtCuNi precursor by arc-melting the pure metallic components (purity > 99.95%) under a Ti-gettered Ar-atmosphere (purity > 99.999%). We then inductively alloyed the elemental P with the PtCuNi precursor in a fused-silica tube under Ar-atmosphere. In order to minimize the oxide content, the alloy was subjected to a fluxing treatment in dehydrated $B_2O_3$ for more than 6 hours at 1473 K. The ribbons were produced by melt spinning of the master alloy on a rotating copper wheel under high-purity Ar-atmosphere. The resulting glass ribbons of $Pt_{42.5}Cu_{27}Ni_{9.5}P_{21}$ at.% had a thickness of 20 μm and a width of 2 mm.

**HPAGs and HPQGs:** We employed a combination of pressure devices, including the Belt press, multi-anvil press, and diamond anvil cell (DAC). To achieve the most controllable and precise compression conditions, we selected nominal pressures of 5 and 10 GPa. The pressure were also chosen to maximize the change in density with respect to the glass bulk modulus of 189 GPa (estimated from the structural evolution under pressure[10] and the connection between the First Sharp Diffraction Peak and density[26]). While higher pressures would be desirable to enhance this density changes, previous reports have shown that $T_g$ increases faster than the crystallization temperature $T_X$[21], limiting the maximum pressure range for glass/liquid stability under both high pressure and high temperature. A rough estimation suggests that this bound is around 15/20 GPa for our composition.

To quench at high pressure from the supercooled liquid phase, we estimated the evolution of the glass transition temperature at 5 GPa. Previous reports based on both in-situ[21] and ex-situ[13,17] measurements indicate a shift of $dT_g/dP \approx 4$ K/GPa in the $Zr_{46.8}Ti_{8.2}Cu_{7.5}Ni_{10}Be_{27.5}$ system, providing a rough estimation for metallic alloys of a pressure induced shift $\Delta T_g(_{5\ \mathrm{GPa}}) \approx 20$ K. Assuming a similar shift for $T_g(P)$ in the $Pt_{42.5}Cu_{27}Ni_{9.5}P_{21}$ system, to ensure to reach the supercooled liquid phase under pressure in the HPQG, the system was heated up to $T_{comp}$=1.05$T_g$(5 GPa). Being the atmospheric glass transition temperature $T_g$(1 atm) = 519 K (measured at 0.33 K/s), this would correspond to $T_{comp}$~$T_g$(1 atm)+50 K.

The HPQG was compressed in a membrane driven DAC using nitrogen as pressure transmitting medium to assure hydrostaticity at high temperature and pressure. Pressure was recorded using the emission spectrum of a ruby sphere, while temperature was measured by a thermocouple in contact with the diamond. Temperature was increase by steps of ~50K from room temperature to 567 K, with constant tuning of the membrane pressure to counter the temperature induced variation of the pressure inside the DAC. The sample was then cooled in isothermal steps to 300 K at 20 K/min and at a constant pressure of 5 GPa. Finally, the pressure was decreased to 1 atm and the resulting glass was recycled for the single-shot FDSC measurements.

For HPAG samples ($T_{\mathrm{comp}} < T_{\mathrm{g}}(P)$), several compressions were performed at the nominal pressures of 5 GPa and 10 GPa, at the same $T_{\mathrm{comp}}$ of $0.98T_{\mathrm{g}}$(5 GPa) K. The 10 GPa compression was performed in a Kawai-type multi-anvil press at the Magmas and Volcanoes Laboratory (LMV, Clermont-Ferrand, France). Ribbons were cut to 1.2 mm long pieces, stacked in a hexagonal boron nitride (hBN) capsule padded with hBN spacers, and surrounded by a graphite furnace. The sample assembly was loaded in a $MgO$-$Cr_2O_3$ octahedron that acts as a pressure-transmitting medium, itself loaded at the center of the 8 tungsten carbides cubes assembly in an 18-11 geometry. Compression to the nominal pressure of 10 GPa was first realized at a rate of $4\times10^{-2}$ GPa/min. The temperature, measured by a Re:W thermocouple placed next to the sample, was increased to the nominal setpoint at 19.4 K/min, followed by an isotherm of 600 s before cooling to ambient temperature. The total duration of the temperature cycle under pressure was 2800 s. Decompression to atmospheric pressure started at $T$ = 323 K. Several 5 GPa compressions were performed in a belt-type apparatus at the X'Press platform of the Néel Institute (Grenoble, France) and the PLECE platform of the Institute of Light and Matter (Villeurbanne, France). Samples were cut to 2 mm long pieces and stacked in an hBN capsule padded with hBN spacers, surrounded by a graphite heater, and using pyrophyllite as the pressure-transmitting medium. The compression to the nominal pressure was first realized at a rate of 0.27 GPa/min (Néel) or 0.1 GPa/min (PLECE). The temperature was calibrated as a function of the power delivered to the graphite furnace, and increased to the nominal setpoint at 20 W/min, which corresponds to a heating rate of 23-30 K/min over the full temperature range, followed by an isotherm of 600 s before cooling to ambient temperature. More information on all compressed glasses can be found in the SM.

**Single-shot FDSC measurement:** To minimize annealing effects and release of the compression upon heating, and to increase the sensitivity of the thermal measurements, we employed FDSC (Mettler Toledo FDSC2+) with a fast-heating rate. This also provides information about the evolution of enthalpy during devitrification, complementing the XPCS data. To improve thermal contact, we used deionized water as the contact agent. After the water evaporates, the small amount of residual salt can securely bind the sample to the chip, similar to the effect of silicone oil. All samples measured 50×50×20 $um^3$ for consistency with the sample recycled after *in-situ* HP XPCS measurements. Heating and cooling rates were controlled between 100-1000 K/s, and measurements were conducted under a protective $N_2$ flow of 100 mL/min. Each sample was used once only, and to ensure repeatability, the measurements were conducted using three identical samples for each heating/cooling rate. To probe if the glass state after densification can be recovered by melting, we applied an *in-situ* melting-quenching protocol in FDSC, where the HPAG was first heated to 923 K for melting, maintained for 0.1 s, then quenched to 300 K at 10,000 K/s, the quenched sample was subsequently measured at a heating rate of 500 K/s (see Fig. S1 for further detail).

To visually highlight the differences in $T_g$ among different samples, we rescaled the data (this does not affect the actual values of $T_g$ and $T_f$). The baseline was obtained by reheating the sample to the crystallization temperature and then scanning again, after which the new data were subtracted to provide the baseline-corrected data. We used the heat capacity change ($\Delta$Cp) at the glass transition and compared it with the $\Delta$Cp from standard DSC scans where the sample mass could be measured. In the standard DSC (20K/min), where the sample mass can be determined, the heat capacity change during the glass transition of the HPAG is the same as that of the ref sample (see Fig. S2). Therefore, we used the height of the heat capacity change measured by FDSC as the standard for mass normalization (by division). The estimated sample mass in FDSC based on the $\Delta$Cp value is between 0.5-1 ug.

**XRD:** The structure of the metallic glass samples was monitored during heating-cooling temperature cycles at atmospheric pressure using XRD performed on the ID15A beamline of the European Synchrotron (ESRF, Grenoble, France)[57], with an incident beam energy of 68.5 keV. Scattered intensity was recorded with a Pilatus3 X

CdTe 2M with two different sample-detector distances: 0.17 and 1.09 m. The longer sample-detector distance was chosen to obtain the highest resolution of scattering vectors corresponding to the FSDP, while the shorter sample-detector distance was used to extract the scattered intensity of scattering vectors of maximum 30 Å$^{-1}$, enabling the computation of the pair distribution functions. Diffraction patterns were azimuthally integrated using routines from the pyFAI library[58], and locally implemented corrections were employed for outlier rejection, background subtraction, polarization of the X-rays and detector geometry, response, and transparency, in order to yield 1D diffraction patterns. *S(q)* was extracted from the coherent scattered intensity $I^c(q)$ as $S(q) = \frac{I^c(q)-|\langle f(q)\rangle|^2}{|\langle f(q)\rangle|^2}$, where $\langle f(q)\rangle = \sum_\alpha c_\alpha f_\alpha(q)$ with $f_\alpha(q)$ and $c_\alpha$ the atomic form factor and atomic concentration for each chemical specie α. The top 50% of the FSDP was reconstructed analytically by fitting the ad-hoc function $S(q) = y_0 + A\frac{1}{1+e^{-\frac{q-q_c+\omega_1/2}{\omega_2}}} \times \left[1 - \frac{1}{1+e^{-\frac{q-q_c-\omega_1/2}{\omega_3}}}\right]$ to the data. The position of the peak was computed as the center of mass of the reconstructed peak, to account for the changing asymmetry. While the 50% threshold is chosen arbitrarily to avoid the contribution of the pre-peak and second peak, the consistency of the results when considering the top 40%, 30% and 20% of the peak was verified. The reduced pair distribution function *G(r)* was obtained by Fourier transform of the structure factor such as $G(r) = \int_0^{+\infty} q(S(q)-1)\sin(qr)dq$. The positive parts $G(r) > 0$ of the pair distribution function were all modelled with the Pearson VII distribution function $G(r) = \frac{A}{\alpha\beta\left(m-\frac{1}{2},\frac{1}{2}\right)}\left[1+\frac{(r-\mu)^2}{\alpha^2}\right]^{-m}$, where $\alpha$, $\mu$ and $m$ are fitting parameters for the width, position, and kurtosis of the peak respectively, while $\beta()$ is the Beta function. Temperature control was performed in a Linkam THMS600 cell, with samples glued on a copper foil in contact with the heating element. The cell was flushed with a constant flux of nitrogen to prevent oxidation of the samples, and the reliability of the sample temperature was tested using the equation of state of a nickel calibrant.

**XPCS:** XPCS experiments were performed on the ESRF's ID10 beamline (Grenoble, France), using an incident beam energy of 21.67 keV ($\Delta E/E=1.4\times10^{-4}$). A partially coherent beam was then focused using a 2D Be lens transfocator to 50.5x14.2 $\mu m^2$ (HxV, FWHM), and cut to 8x8 $\mu m^2$ by guard slits. Scattering patterns were recorded with an Eiger2 4M CdTe detector located 5 m downstream from the sample position, at a scattering vector corresponding to the maximum of the FSDP, i.e. (2.873±0.006) $Å^{-1}$ at zero pressure and (2.898±0.001) $Å^{-1}$ at 5 GPa. The $g_2(q,t)$ intensity-intensity correlation functions were extracted from all acquired frames using the sparse correlator algorithm described in reference[59]. The ISF was retrieved from the $g_2(q,t)$ function using the Siegert relation $g_2(q,t) = 1 + \gamma.|F(q,t)|^2$, whose application to non-ergodic systems is ensured by the many q-equivalent speckles in large area detector. The time resolved evolution of the dynamics within a single XPCS scan was assessed from the Two-Times Correlation Function (TTCF), which is obtained from the evaluation of the correlation values $C(q,t_1,t_2)$ of each pair of scattering patterns acquired at times $t = t_1$ and $t = t_2$. From the cross-correlation figure such as these displayed in Fig. 5, horizontal (lower quadrant) and vertical (upper quadrant) lines correspond to $g_2(q,t)$ functions with respect to the reference taken at $t = t_1 = t_2$. The equilibration in the supercooled liquid phase is deduced from the shape of the ISF, the liquid state being reached when the $\beta$ parameter inferred from a KWW fit decreases below 1[41]. Another sign that the liquid state has been reached is the temperature dependence of the relaxation time, which follows the macroscopic Vogel-Fulcher-Tamman (VFT) law[60], and the overlap of data acquired on cooling and heating the liquid.

**Acknowledgments**

We gratefully acknowledge the ESRF (Grenoble, France) for providing beamtime, including experiments carried out on ID10 and ID15A beamlines under the proposal HC4986 and LTP project HC4529. The Partnership for Soft Condensed Matter (PSCM) at the ESRF is also acknowledged for providing support facilities for sample characterization. We thank J-L. Barrat and K. Martens for the interesting scientific discussions, K. Lhoste and D. Duran for assistance on the ID10 and ID15A beamlines, and J. Jacobs for help with the high-pressure experiments. The multi-anvil apparatus of Magmas and Volcanoes Laboratory is financially supported by the French National Centre for Scientific Research (National Instrument of the National Institute for Universe Sciences).

**Funding:**

This project received funding from the European Research Council (ERC) under the European Union's Horizon 2020 research and innovation program (Grant Agreement No 948780).

**Author contributions:**

AC, JS, AR, SL, NC, MF, EP, TD, YC, FZ, MC and BR conducted the XPCS experiments. AC, JS, AR, BR, GV and MdM carried out the XRD experiments. GG assisted the HP-XPCS measurements. AC and GV analyzed the XRD/PDF data, and AR, SL, BR and AC analyzed the XPCS data. JS performed the FDSC and DSC measurements and analyzed the resulting data with the support of DC. AC, JS, AR and SL conducted all the *ex-situ* compressions. SLF (PLECE platform) and ML and CG (X'Press platform) participated in the belt press densifications, while GL helped in the multi-anvil densifications. AC, JS and BR wrote the manuscript with contributions from all authors. AC and JS contributed equally to this work. BR conceived the work and led the investigation.

**Competing interests:**

The authors declare that there are no financial or non-financial competing interests.

## Data and materials availability:

The study generated few TB of data, which are available from the corresponding authors upon reasonable request.

## Figures and Tables

| Sample | $P_{comp}$ | $T_{comp}$ or $T_a$ | $T_{g,onset}$ (K) | $T_f$ (K) | $\frac{dT_g}{dP}$ |
|---|---|---|---|---|---|
| Reference | 1 atm | 1.05 $T_{g,1atm}$ | 561 | 507 | -- |
| HPQG | 5 GPa | 1.05 $T_{g,5\ GPa}$ | 573 | 497 | $(\frac{dT_g}{dP})_{ex\text{-}situ} > 0$ |
| HPAG | 5 GPa | 0.98 $T_{g,5\ GPa}$ | 552 | 537 | $(\frac{dT_g}{dP})_{ex\text{-}situ} < 0$ |

**Table 1. Summary of the FDSC results:** Thermal and pressure protocol, $T_{g,onset}$, $T_f$, and evolution of $T_g$ with pressure, for the reference glass at 1 atm, and the high pressure compressed glasses at 5 GPa obtained by quenching from the liquid (HPQG) or by annealing under pressure in the glass (HPAG).

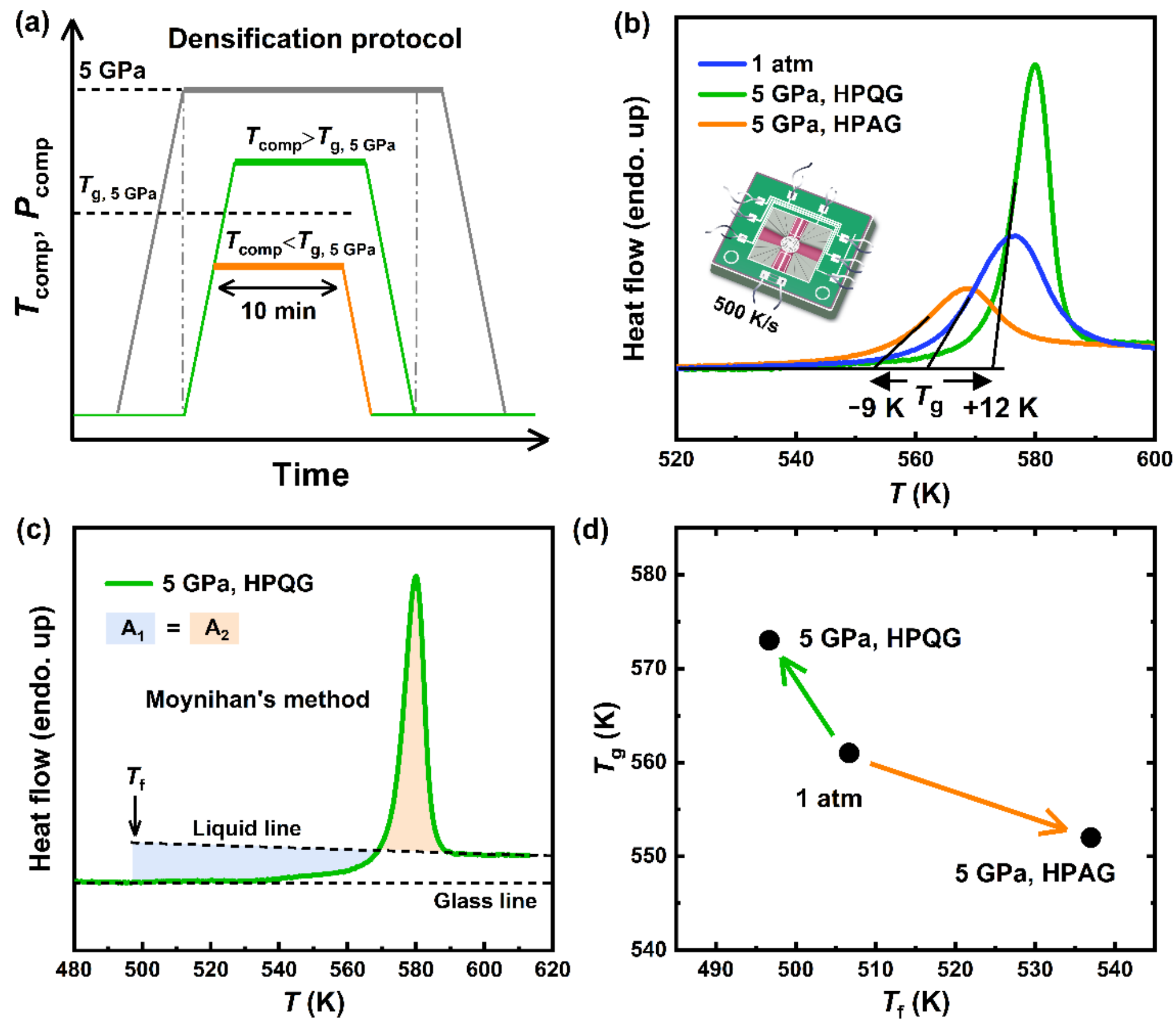

**Figure 1. Thermal response of compressed glasses.** (**a**) Sketch of the *ex-situ* densification protocol. (**b**) FDSC curves of glasses obtained with different thermal and pressure protocols on samples pre-cooled at 20 K/min from their high temperature state and then heated with the FDSC at 500 K/s. The green line is the HPQG produced by cooling at 5 GPa from the high-density liquid. The orange line is a HPAG compressed at 5 GPa in the glass state, and the blue line represents a reference glass obtained using the same thermal protocol but at 1 atm. The intersection points of the black lines represent the $T_g$ (defined here as the onset of the glass transition region), and the arrows signal the relative shift with respect to the reference. (**c**) Evaluation of the fictive temperature ($T_f$) using the equal-area construction for the 5 GPa HPQG (other glasses in Fig. S3). The 'liquid' and 'glass' lines represent the linear fits of the first heat flow curves in the supercooled liquid and glass, respectively. (**d**) Corresponding $T_f$ and $T_g$ obtained from the FDSC data in panel (**b**).

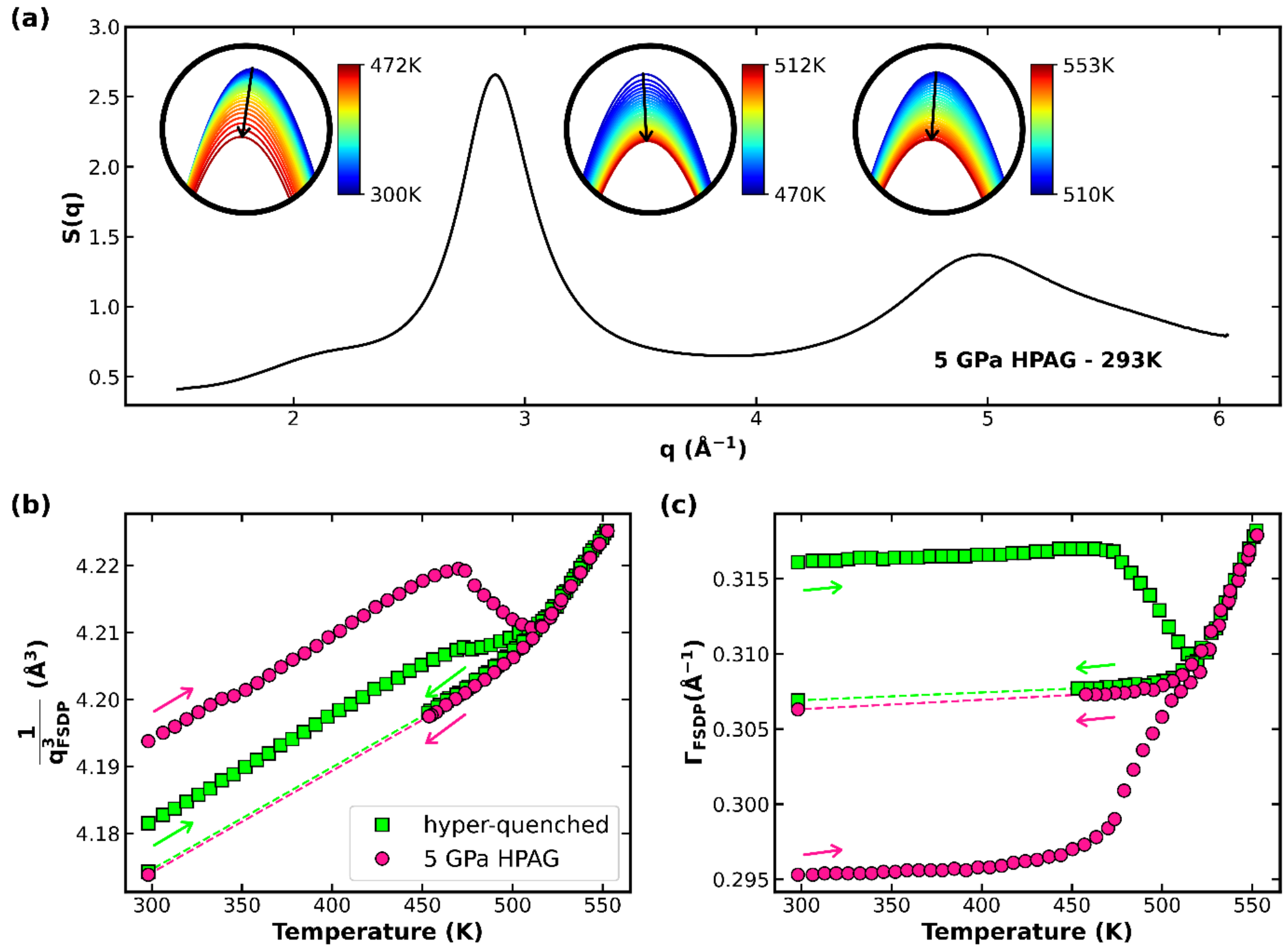

**Figure 2. Apparent structural rejuvenation.** (**a**) Static structure factor, $S(q)$, measured with XRD at room temperature in the HPAG compressed at 5 GPa. The insets show the evolution of the intensity of the first sharp diffraction peak (FSDP) on heating. (**b**) and (**c**) Temperature dependence of the corresponding FSDP center of mass, $q_{\mathrm{FSDP}}$ (**b**) and full width half maximum (**c**) on heating and cooling beyond $T_g$.(pink circles). The data are compared to those of a hyper-quenched glass heated at 1 atm to emphasize the apparent structural expansion in the compressed glass and the increased order in the HPAG.

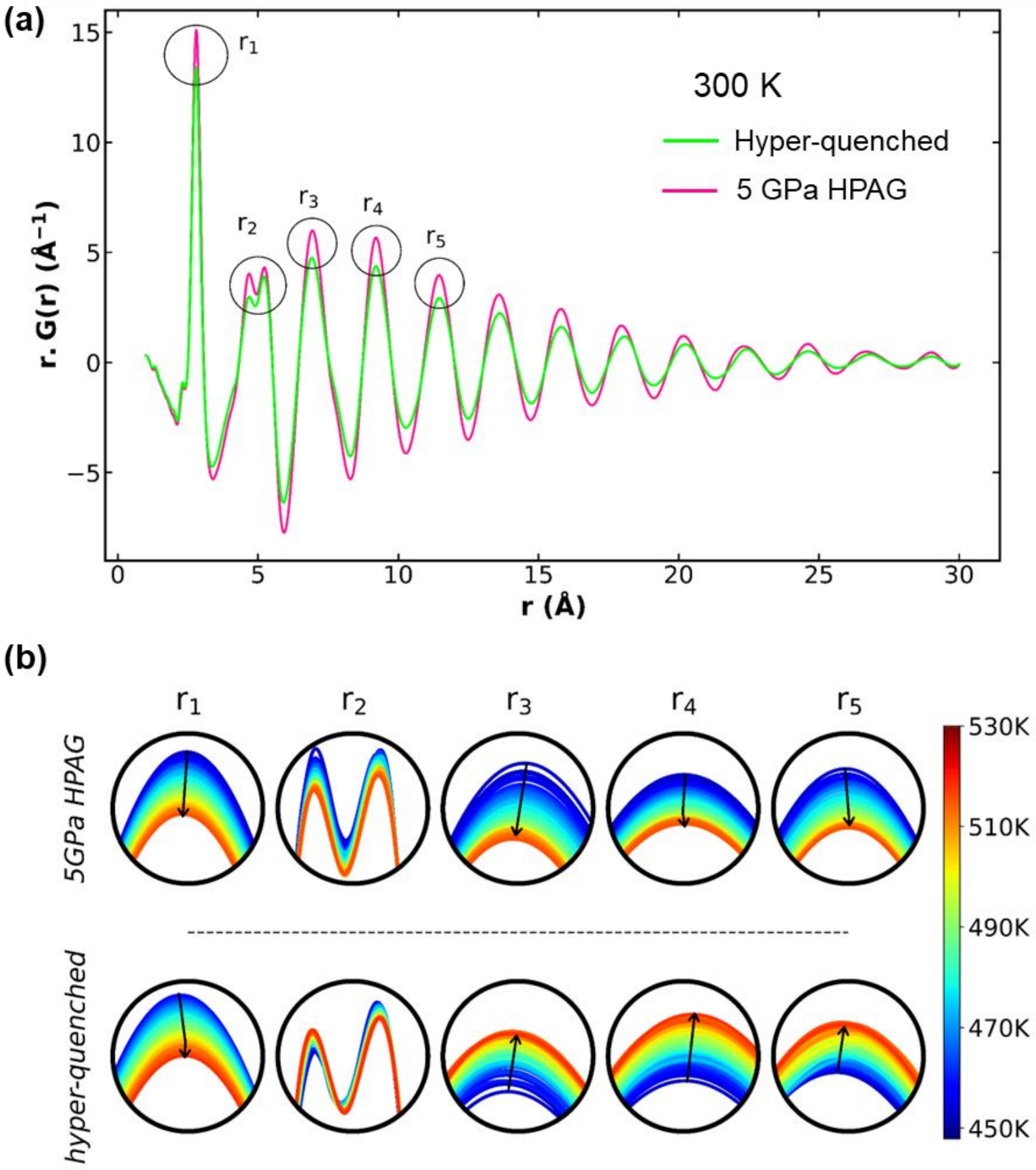

**Figure 3. Real Space analysis.** (**a**) Pair distribution function multiplied by the distance *r* of the neighboring shells obtained in a second XRD experiment on heating hyper-quenched and compressed (5 GPa HPAG) glasses in the glass transition region. (**b**) The insets show the evolution of the maximum peak intensity with respect to temperature of the first five shells in both glasses upon heating up to $T_g$.

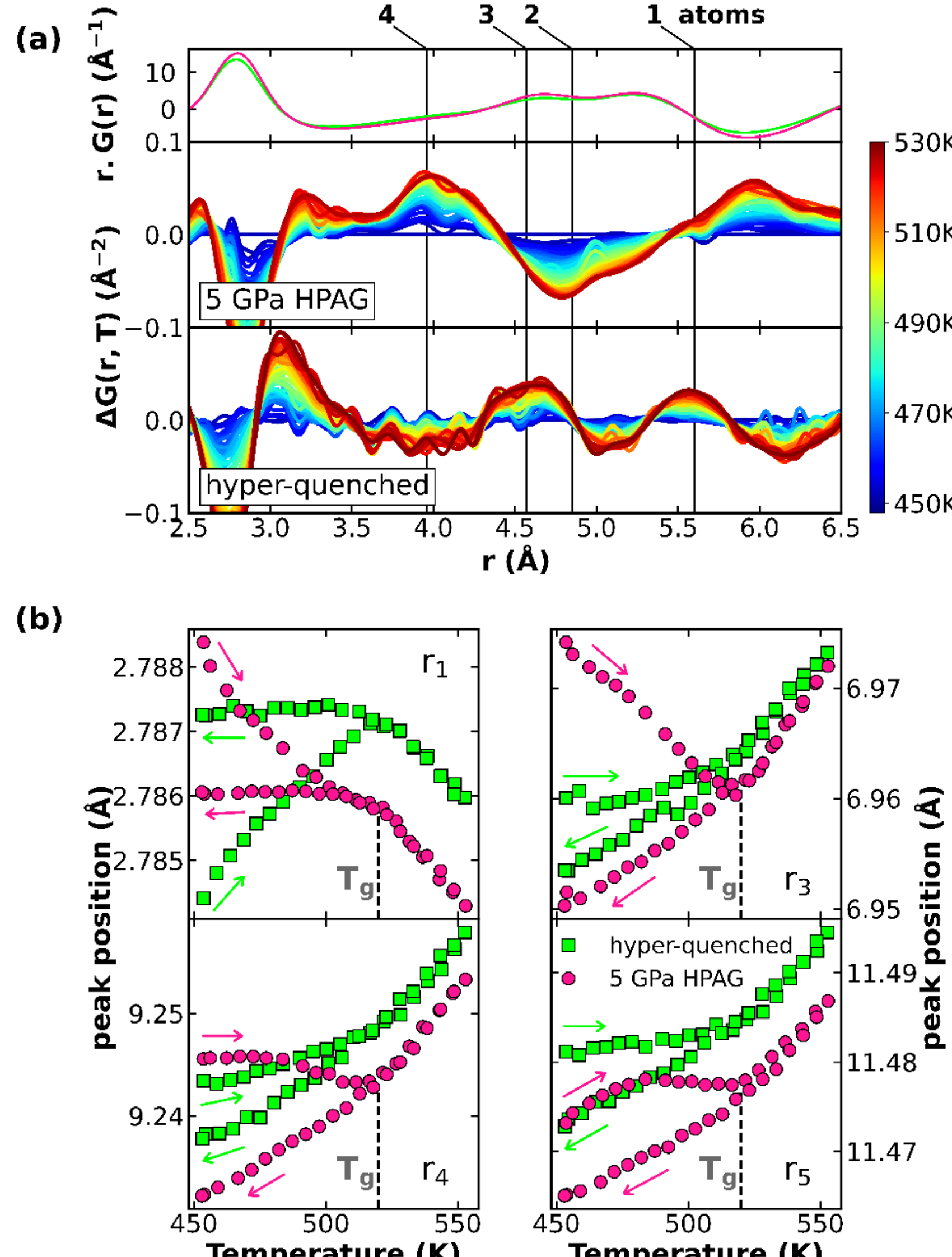

**Figure 4. Anomalous thermal expansion and absence of structural recovery in thermally rejuvenated glasses.** (**a**) Differential pair distribution function $\Delta G(r,T) = G(r,T) - G(r,T = 453K)$ for both the hyper-quenched and HPAG samples around the second coordination shell upon heating up to $T_g$. Vertical lines indicate most probable distances for cluster connection through 1,2,3 and 4 atoms. (**b**) Evolution of the distance of four representatives neighboring shells covering the short and medium range order (SRO and MRO) for both the HPAG and hyper-quenched glass, upon heating and cooling (see Supplementary Materials for the other shells). The dashed lines indicate the $T_g$ of the HPAG, while the colored arrows indicate data acquired during heating and cooling.

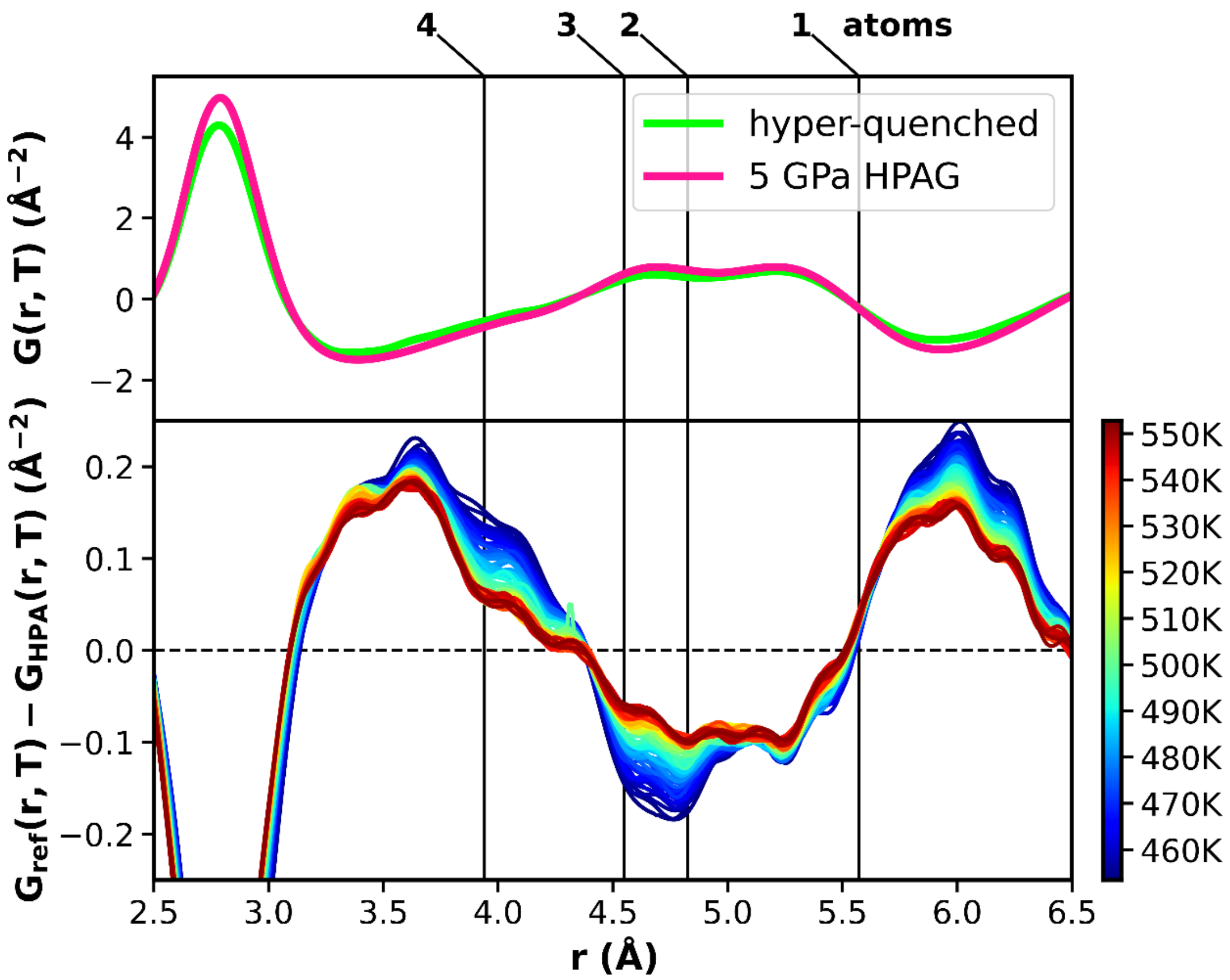

**Figure 5. Structural differences between As-cast and HPA samples.** Room temperature pair distribution functions of hyper-quenched and HPAG samples, and the respective cross-samples differential distribution function as a function of temperature. Negative values correspond to dominant features in the HPAG.

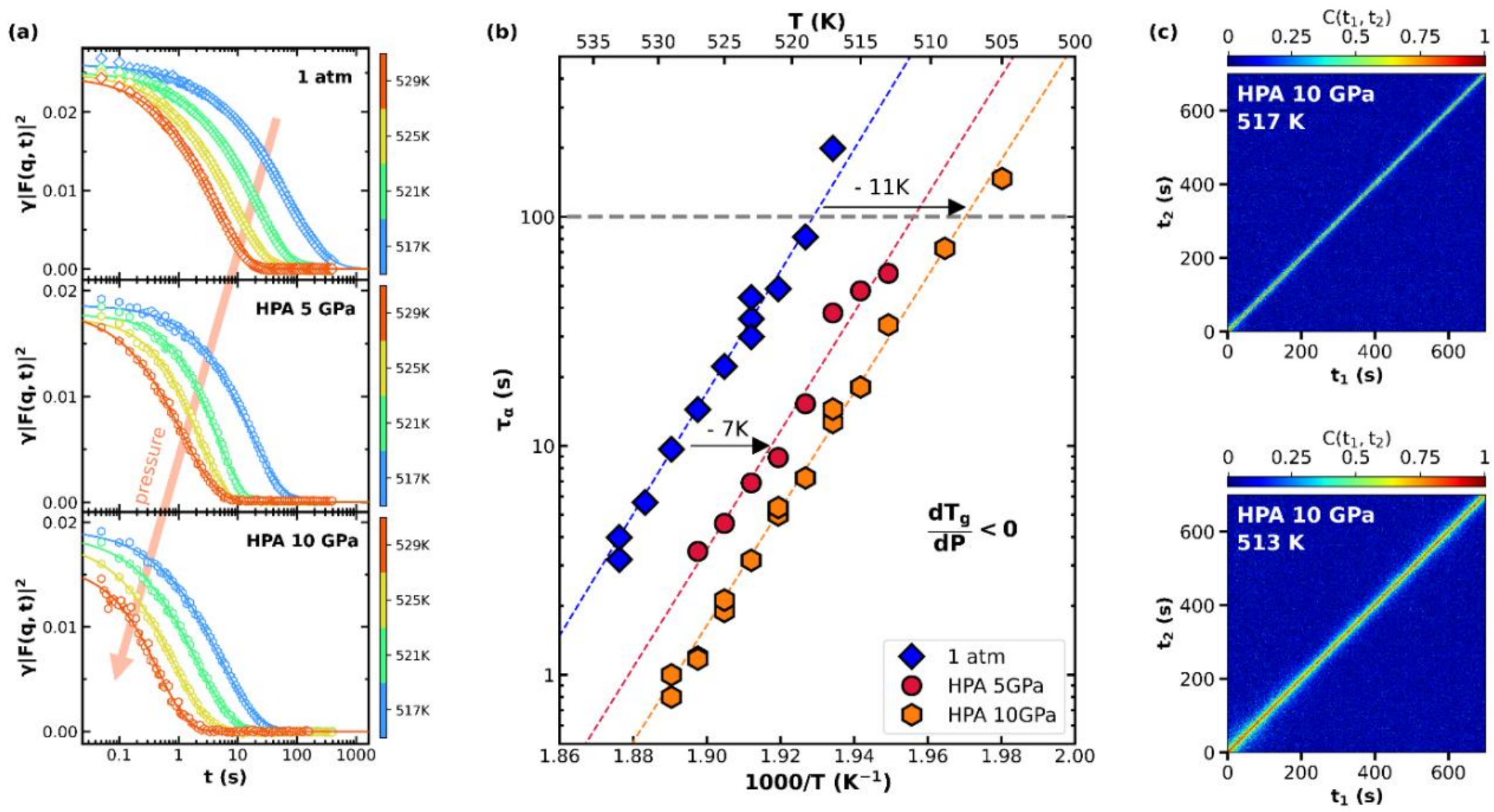

**Figure 6. Absence of dynamic recovery.** (**a**) Temperature dependence of ISFs measured using XPCS in the supercooled liquid phases obtained after heating 5 GPa and 10 GPa HPAGs above their $T_g$, together with a reference liquid at 1 atm. The arrow indicates accelerating dynamics with increasing pressure for the *ex-situ* compressed samples. Solid lines are fits of the KWW function. (**b**) Corresponding $\tau_\alpha$. Dashed lines are Arrhenius fits. Uncertainties are within the size of the symbols. The arrow indicates the lower $T_g(P)$ of the HPAGs following the convention $T_g=T(\tau_\alpha=100\text{ s})$, which gives $T_g(5\text{GPa})_{HPA}=511$ K and $T_g(10\text{GPa})_{HPA}=507$ K. (**c**) Two-times correlation functions measured using XPCS showing stationary dynamics. These functions are a time-resolved evolution of the ISF and describe the correlation level $C(q,t_1,t_2)$ between two frames acquired at times $t_1$ and $t_2$.

# Supplementary Material:

## Break-down of the relationship between α-relaxation and equilibration in hydrostatically compressed metallic glasses

Antoine Cornet*, Jie Shen* *et al.*

*Corresponding authors. Email : antoine.cornet@neel.cnrs.fr, jie.shen@neel.cnrs.fr, beatrice.ruta@neel.cnrs.fr

### 1. Summary of all samples reported in the work and the employed experimental approach

As discussed in the main manuscript, the complexity of the glassy state is highly related to the existence of a multitude of possible metastable states which can be reached by following different thermo-mechanical paths, and by the dependence of several properties of glasses on the applied protocol used to measure it, as for instance for the glass transition temperature. As a consequence, depending on the technical requirements and scientific goals of each technique, we find different $T_g$ values, or we have used additional samples. To help the reader, we report here a brief summary of the samples and results. As discussed in the manuscript, all data provide a unique picture of the pressure response of MGs, although we have used different techniques and compared different protocols and glasses.

**List of samples:**

In our work, we study the response of $Pt_{42.5}Cu_{27}Ni_{9.5}P_{21}$ MGs to thermo-mechanical treatments. For this goal, we used glasses prepared following different protocols.

1) High pressure quenched glasses (called "HPQG"): glasses compressed at 5 GPa and kept at $1.05T_g$(P=5GPa), cooled at 20 K/min and decompressed.

2) High pressure annealed glasses (called “HPAG”): glasses compressed at 5 GPa and kept at $0.98T_g$(P=5GPa) for 10 min, cooled at 20 K/min and decompressed.

3) Reference glass (called “1 atm”): glasses heated at $1.05T_g$(P=1 atm), and then cooled at 20 K/min.

4) As-cast glass (called “hyper-quenched”): glasses prepared by quenching the melt with $10^6$ K/s.

5) To confirm the existence of multiple liquids and study the effect of increasing compression pressure, a second HPAG has been prepared with the same protocol as #2 but keeping the pressure at 10 GPa.

**FDSC results:** Table 1 in the main manuscript reports the thermal and pressure protocol, $T_{g,onset}$ and $T_f$ for the reference glass at 1 atm, and the high pressure compressed glasses at 5 GPa from the supercooled liquid (HPQG) and from the glass (HPAG). All samples have been previously cooled from the compression temperature, $T_{comp}$, or annealing temperature $T_a$ for the reference, with a nominal rate of 20 K/min and subsequently measured with FDSC with a rate of 500 K/s. The larger values of the glass transition temperature with respect to those obtained with the other techniques are a simple consequence of the fast heating rate used during the scans. As the heating rate used for FDSC is way larger than the cooling rate, this naturally explains why $T_{onset}$ is larger than $T_f$. If the heating and cooling rates were the same, $T_f$ (approximately equal to the mid-step $T_g$) would be slightly different than $T_{g,onset}$, due the presence of non-$\alpha$ contributions in the transformation from glass into liquid[1].

**XRD and PDF results:** To understand the structural changes associated to the surprising lower thermal stability of HPAGs with respect to glasses annealed at 1 atm, in Figure 2 and Figure 3 of the main manuscript we have compared the HPAG compressed at 5 GPa used also for FDSC measurements with both a hyper-quenched glass (i.e. obtained with a fast cooling rate of $10^6$ K/s from the melt) and the standard reference at 1 atm (which can be simply retrieved by heating and subsequently cooling the hyper-quenched glass above $T_g$). Being the hyper-quenched glass the archetype of a structurally rejuvenated glass with extra free volume, the comparison between the two systems allows to get more insights into the "apparent" volume expansion of the HPAG. The analysis of the PDF results allows to clarify the structural differences between as cast and high pressure annealed glasses at the level of the different coordination shells, and to study the devitrification mechanism.

**XPCS results:** XPCS measurements were used to corroborate the XRD results and in particular the absence of equilibrium recovery in the supercooled liquid phase. As dynamical quantities are more sensitive that static observables, the presence of distinct states becomes unambiguous in the dynamical data.

## 2. Kinetic response of HPAGs thermally cycled across $T_g$

The kinetic rejuvenation of HPAGs with respect to glasses annealed in the absence of pressure is independent of the applied heating rate in the 100-1000 K/s range (Figure S1), and corresponds to a shift of -9.75±0.5 K at 5 GPa, i.e. ~20 K below the $T_g$ of the HPQG obtained by compressing the dense liquid at the same pressure and releasing the pressure in the glassy state.

FDSC scan can also detect that the thermo-mechanical history of the HPAG cannot be fully erased if the system is heated above $T_g$. As shown in Figure S1b $T_g$ values obtained during a second heating run (Run2 in the inset) do not overlap with reference values, independently of the applied heating rate. The reproducibility of the FDSC data is confirmed by the superposition of the curves measured on three different HPAGs at 5 GPa (Figure S1c). Even if the difference between HPAGs and references decreases due to the major compression released within the first heating run, a $T_g$-shift is still visible and reproducible even after successive thermal cycling (Figure S1d). The HPAG reverts to the original glass only if quenched from above the melting temperature (Figure S1d). The reproducibility of the FDSC data is confirmed by the superposition of the curves measured on three different HPAGs at 5 GPa (Figure S1e).

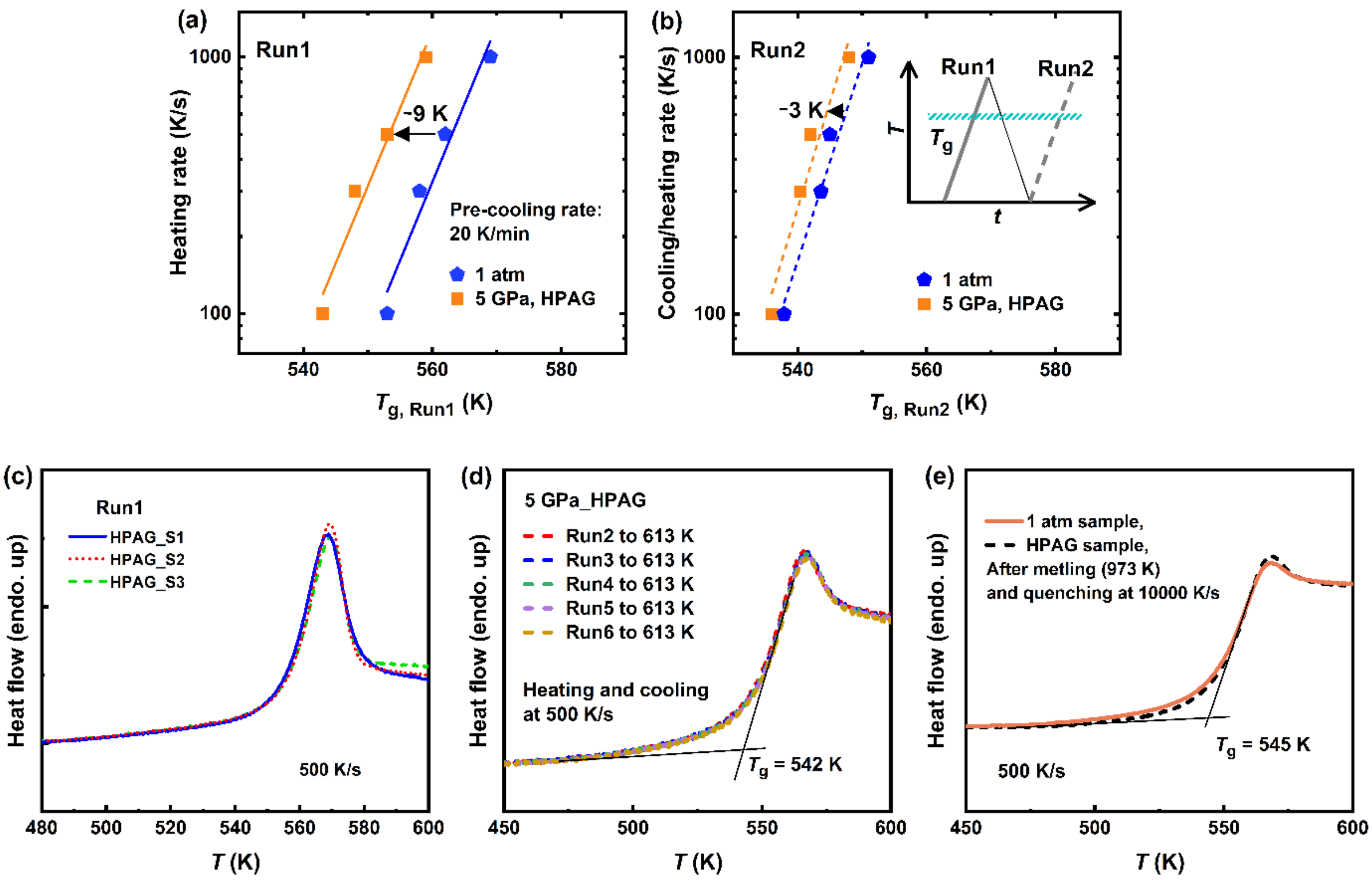

**Figure S1.** a) Heating rate dependence of $T_g$ measured with FDSC for a HPAG compressed at 5 GPa and $0.98T_g$(5 GPa), and a reference glass obtained with the same protocol but at 1 atm (blue line). The data are obtained during a first heating run up to 613K, and (b) after cooling the samples to 300 K and reheating them to 613 K (Run2, panel b). Solid and dashed lines in (a,b) are linear fitting and different points correspond to independent data acquired on different samples. (c) Reproducibility of the measured for the HPAGs. (d) The Run 2 protocol was repeated for a total of 6 cycles in FDSC and only Runs 2 to 6 are shown here. The well overlap of the 5 curves indicates that after 5 glass-liquid transitions, the new amorphous state still does not return back to the reference glass state. e) Comparison between the reference state and a 5 GPa HPAG after undergoing a fully melting and quenching process. The agreement between the two curves shows that the thermal changes induced by the compression can be totally erased only after melting the sample.

## 3. Thermal properties of HPAGs using standard differential scanning calorimetry (DSC)

Under a heating rate of 500 K/s in FDSC, the sample consistently exhibits an overshoot in the heat flow above $T_g$, which is due to the fact that the heating rate applied to the sample is much higher than the cooling rate used during its preparation (20 K/min).

Figure S2 shows the heat flow profile of a HPAG compressed at 7 GPa and 0.98$T_g$(5 GPA) subsequently heated at 20 K/min with standard DSC and the comparison with a reference sample subjected to the same thermal protocol but at 1 atm. The HPAG exhibits a noticeable exothermic undershoot before $T_g$ and a smaller endothermic peak after $T_g$, indicating a higher-energy state, which aligns with the information from FDSC. We note that in standard DSC the $T_g$ of the HPAG is almost invariant and decreases by only 2K with respect to FDSC, where the $T_g$ of the HPAG decreased by 9 K. This effect is a consequence of the lower sensitivity of standard DSC and to the major pressure releasing process (corresponding to the exothermic peak in DSC) in the HPAGs which occurs on heating with slow rates.

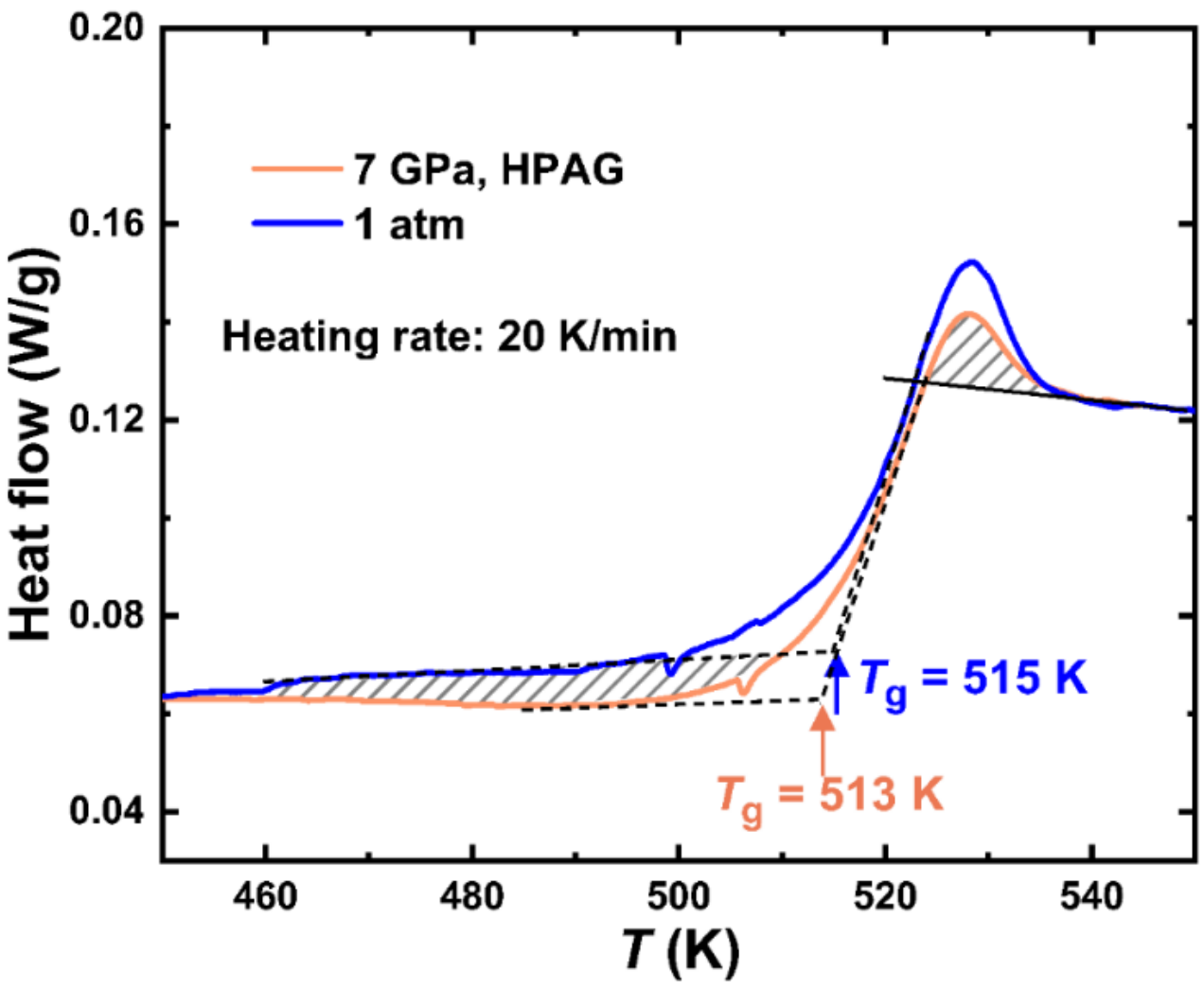

**Figure S2.** DSC curves (Perkin-Elmer DSC 8000) of a HPAG compressed at 7 GPa and 0.98$T_g$(5 GPa) and a reference sample obtained with the same protocol but at 1 atm. The shaded areas represent the exothermic process during the heating and the endothermic peak after glass transition for the HPAG, indicating a higher energy state. In the DSC, the $T_g$ of the two samples does not show a significant difference as in FDSC (Figure 2a) with rapid heating. In FDSC, the rapid heating prevents the HPAG from releasing the compression effect (corresponding to the exothermic peak in DSC) until the glass transition occurs, thereby allowing further insights into the compressed state.

## 4. Evaluation of the fictive temperature

We utilize the heat flow curve from FDSC to estimate the $T_f$ for different samples based on the area-matching method[2,3] following the equation:

$$\int_{T_f}^{T_1 \gg T_g}\left(C_{p,\ \text{liquid}} - C_{p,\ \text{glass}}\right)dT = \int_{T_2 \ll T_g}^{T_1 \gg T_g}\left(C_p - C_{p,\ \text{glass}}\right)dT \qquad (1)$$

Here, $C_{p,\ \text{liquid}}$ and $C_{p,\ \text{glass}}$ represent the specific heat capacities of the liquid and glass, respectively. The enthalpy area matching is shown in Figure S3 and the values in table S2.

As discussed in the main text (Figure 1d), an estimation of the corresponding effective cooling rates of the reference glass can be obtained from the intersection of the $T_f$ with a Vogel-Fulcher-Tamman (VFT) equation with fragility $D^*$=15.3 as previously reported[4]. The reference glass has $T_f$(1 atm)=507 K corresponding to an effective cooling rate of 0.27 K/s in close agreement with the actual rate of 0.33 K/s, validating the procedure. Following the XRD and XPCS results, due to the existence of different liquid states for the compressed glasses we cannot use the literature viscosity data at 1 atm. to estimate the corresponding effective cooling rates.

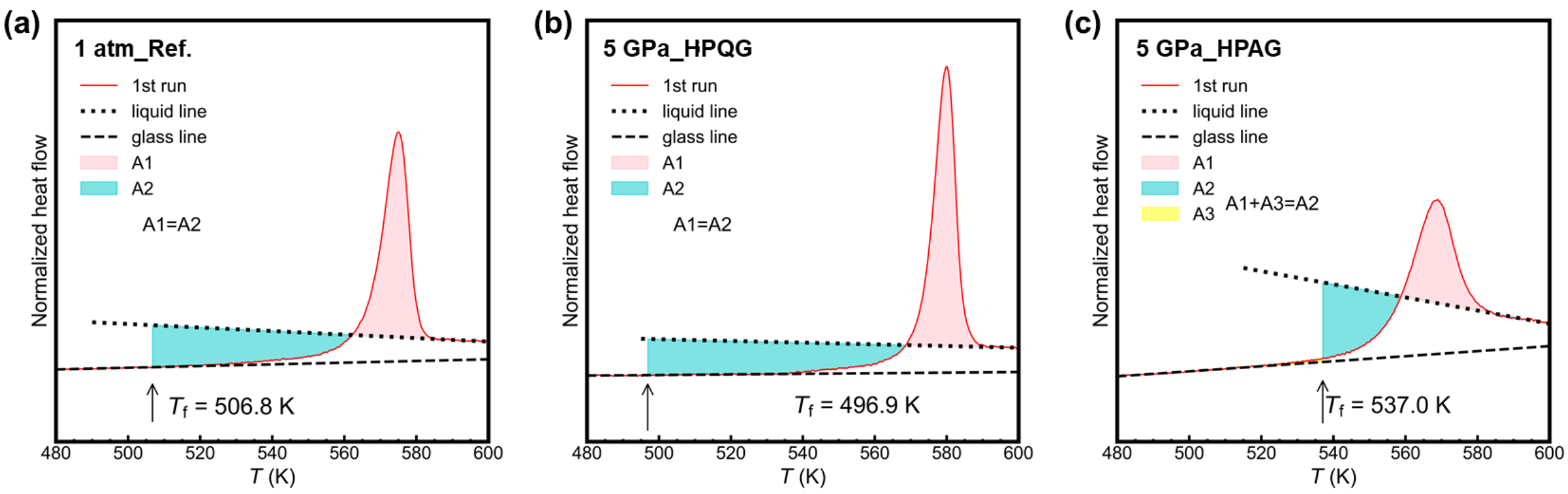

**Figure S3. Evaluation of the fictive temperature ($T_f$) using the equal-area construction.** The 3 panels show the method of calculating $T_f$ for all samples reported in Figure 1 in the main text. The 'liquid line' and 'glass line' represent the linear fits of the first heat flow curves in the supercooled liquid and glassy region, respectively.

## 5. Complete structural evolution of HPAG and as-cast samples

Figure S4 shows the evolution of the intensity of both the *S(q)* data with high *q*-resolution around the first principal diffraction peak and the pair distribution functions upon heating measured in a different experiment, both for the HPAG 5GPa sample.

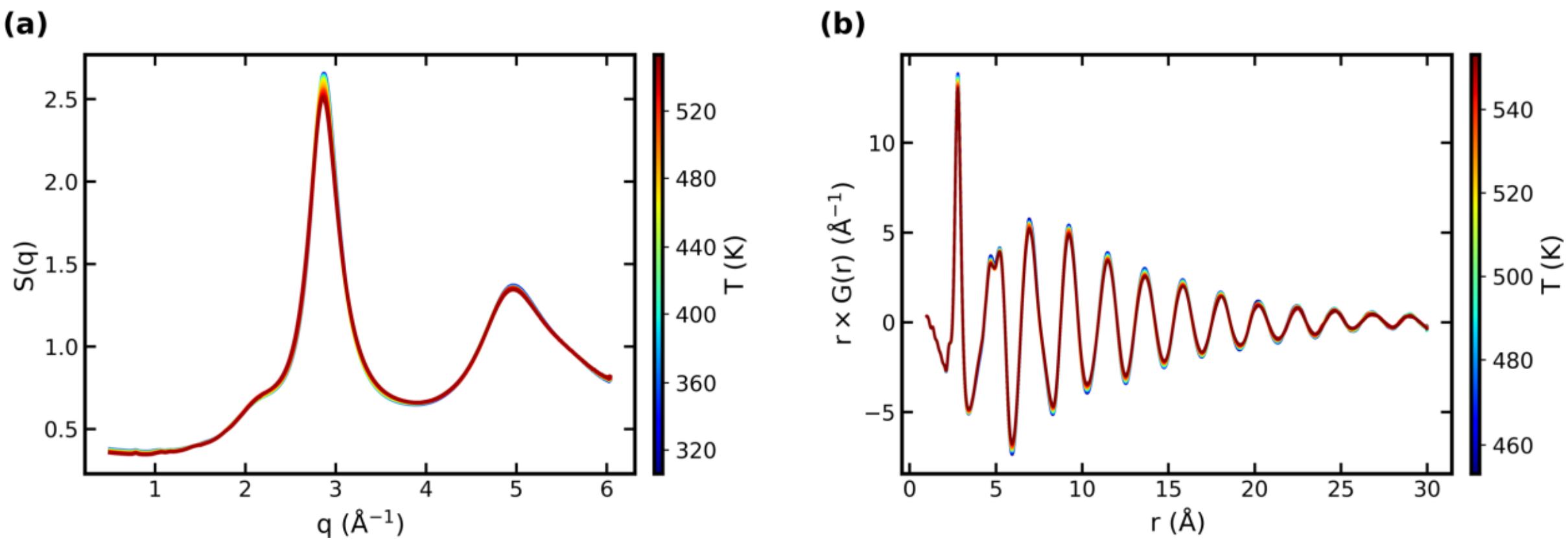

**Figure S4.** High-resolution structure factor S(q) (a) and pair distribution function r×G(r) (b) obtained upon heating the HPAG 5GPa sample at atmospheric pressure. The temperature of each curve is marked from the color coding adjacent to each panel.

From the high $q$-resolution diffraction data, we obtained differential structure factors $S(q,T) - S(q,T = 300K)$ which provide the $q$-dependent structural evolution, as shown in the Figure S5 for the as-cast and HPAG 5GPa samples upon heating.

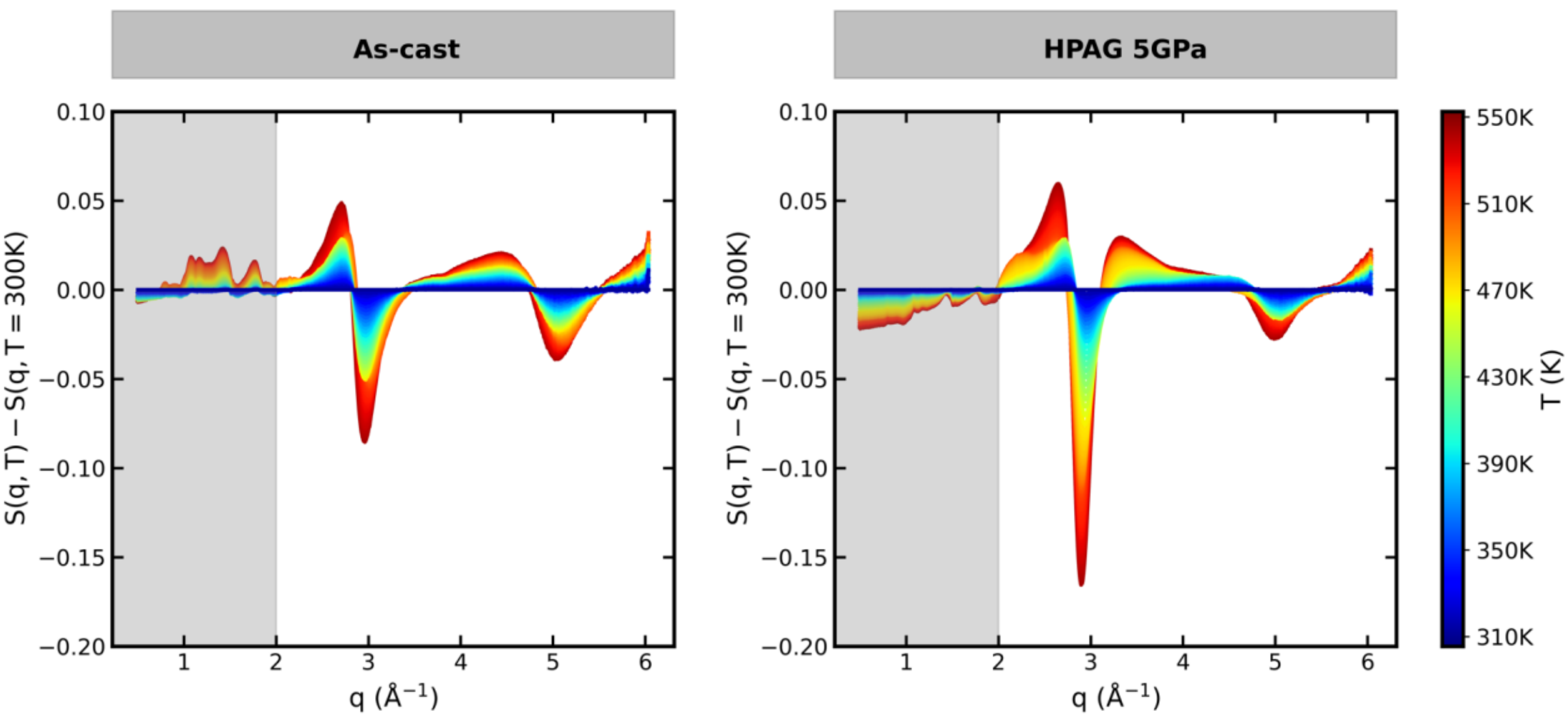

**Figure S5. Differential structure factor pair distribution function upon heating.** Color coded evolution of the intensity relatively to the structure factor S(q) obtained at room temperature for both the as-cast and the HPAG 5GPa samples. The grey shaded area is dominated by artefacts, and does not correspond to the structural evolution of the sample.

At the level of the principal diffraction peak, around 2.9 $Å^{-1}$, the differential plot reveals a shift toward lower $q$ values for the as-cast sample, as indicated by the symmetric pattern around the isosbestic point at $q = 2.9$ Å$^{-1}$, which confirms the release of extra free volume on heating. For the HPAG 5GPa sample however, the positive $\Delta S(q)$ on both side of the main negative feature reveals a widening of the diffraction peak, in addition to a shift to lower $q$, visible through the relative intensities of the two positive features at $q = 2.7$ Å$^{-1}$ and $q = 3.4$ Å$^{-1}$. This widening on heating indicates a diminution of the ordering of the glassy network once the constraints induced by the pressure protocols are partially removed, in agreement with the results of the fits of the FWHM of the *S(q)*, which show as well that the HPAG is more homogenous and ordered with respect to the as-cast glass, and that the system becomes less homogenous on heating through the glass transition (Figure 2c). The same vertical scale used for the two plots reveal that the structural change upon heating is roughly twice as important in the HPAG 5GPa sample relatively to the reference as-cast sample.

The relative ordering/disordering of both samples is also visible from the differential pair distribution function $\Delta G(r) = (G(r,t) - G(r, T_{reference}))$. This time the reference dataset is chosen to be just below the equilibration (or the release of the majority of the pressure induced structural changes) at $T_{reference} = 453K$. The temperature evolution of $\Delta G(r)$ upon heating is shown in the figure S6 for both samples.

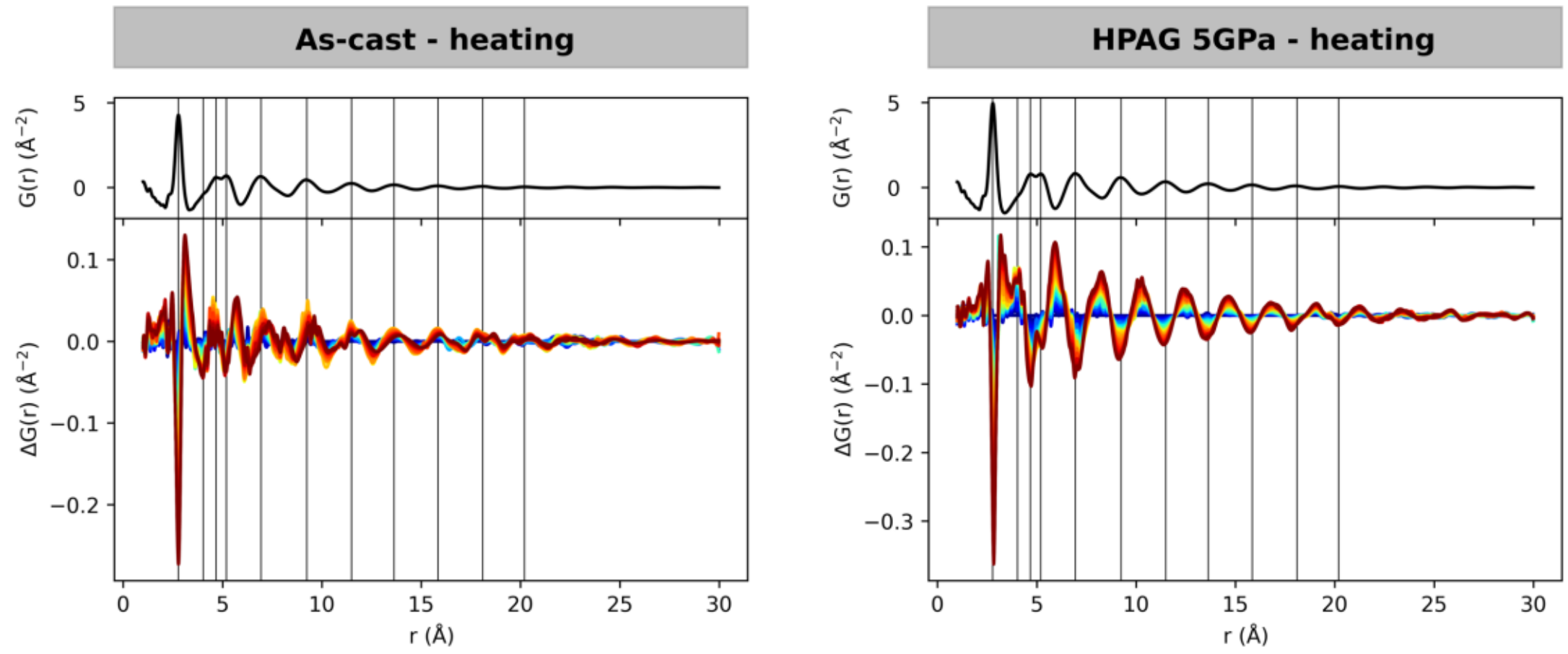

**Figure S6. Differential pair distribution function upon heating.** As-cast and HPAG 5 GPa samples show opposite intensity trend at large length scale upon heating.

The intensity change correlate with the position of the coordination shells on the as-cast sample, while they anti-correlate in the HPAG 5GPa sample, revealing ordering enhancement and reduction respectively. This picture also reveals that, intensity wise, the evolution of the 4th and 5th coordination shells is representative of all the high coordination shells. The cooling behaviour of $\Delta G(r)$ is shown on the Figure S7, using the high temperature data as reference, that is $T_{reference} = 553K$.

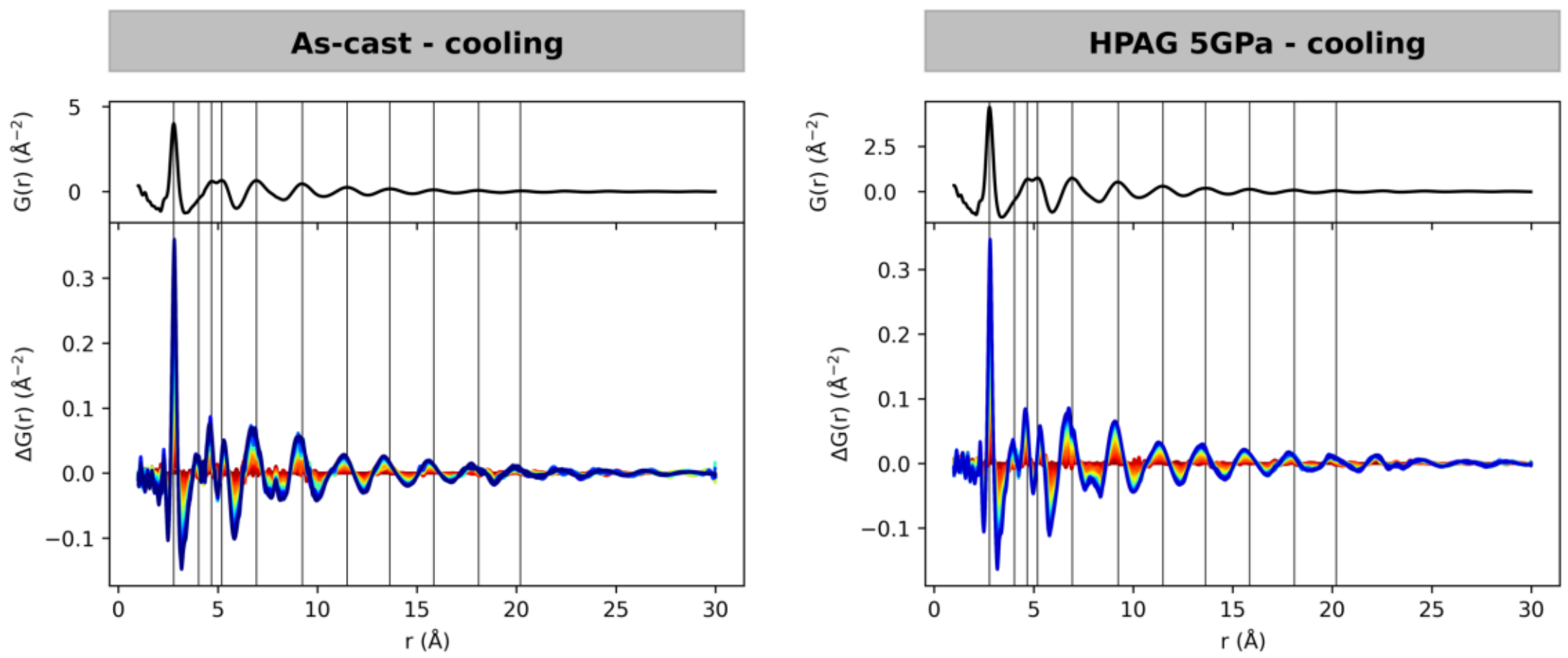

**Figure S7. Differential pair distribution function on cooling.** As-cast and HPAG 5 GPa samples show similar intensity trend at large length scale upon heating.

The relative structural changes on cooling are similar for both samples on cooling. This evolution describes an ordering process, as evidenced by the positive correlation between the intensity change and the position of each coordination shells. It demonstrates that the difference between the two systems that persists upon heating above the glass transition temperature $T_G$ is not affected by the cooling process, including upon crossing $T_G$ on cooling.

The main text features the intensity plot of each coordination shell up to the 5th one only. This is because this 5th coordination shell is representative of the behaviour of all larger coordination shells. This was demonstrated already on the intensity evolution on the Figure S6, and is confirmed from the position evolution on the Figure S8.

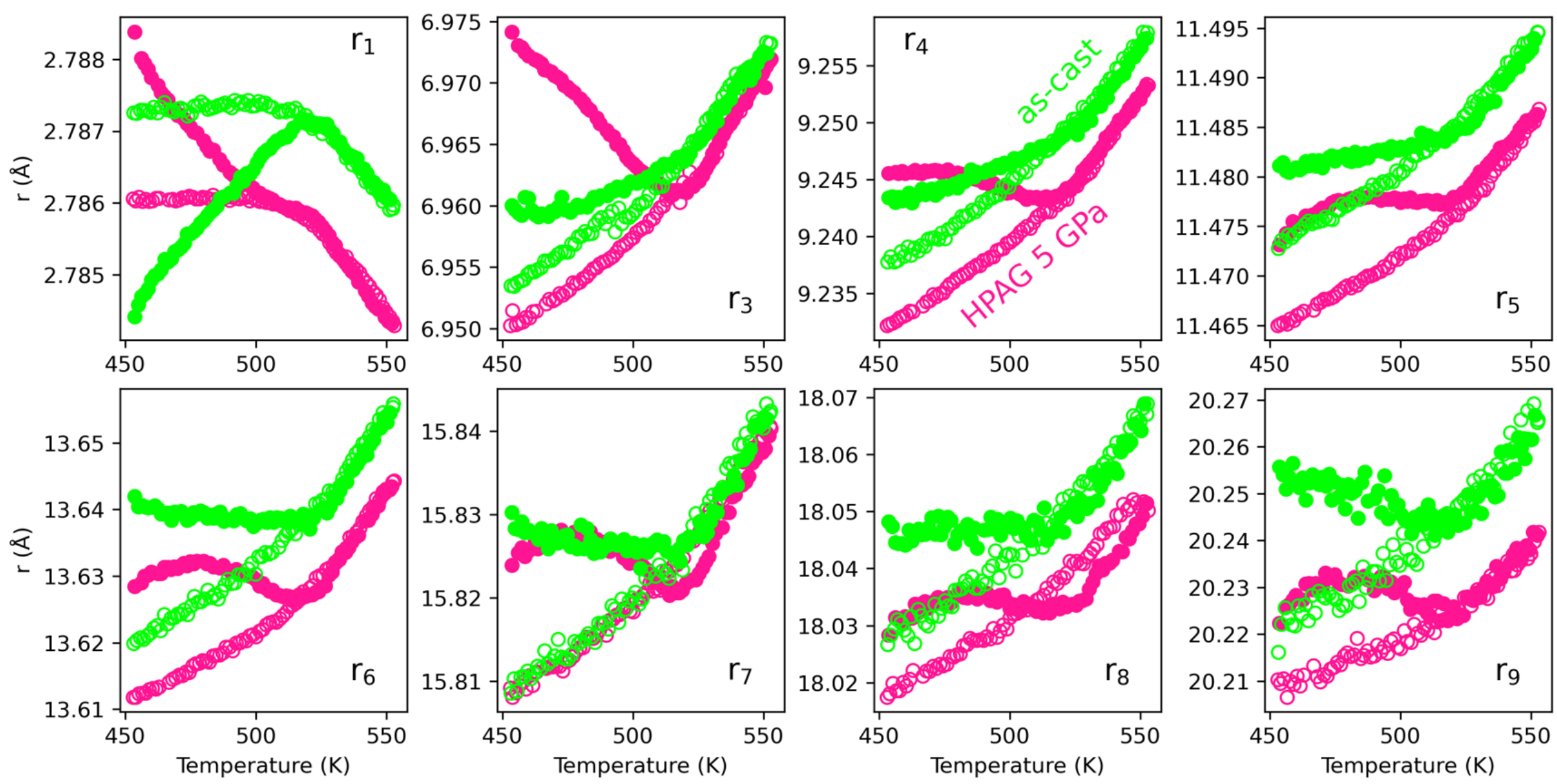

**Figure S8. Temperature evolution of coordination shells.** Plain symbols correspond to data obtained upon heating, empty symbol correspond to data obtain upon cooling.

## <u>5. Pressure dependence in HPAGs</u>

When heated above $T_g$ at ambient pressure, the HPAG transforms into a new system compared to the reference, with faster decay times in the ISFs (see main text, Figure 6a), corresponding to lower relaxation times (see Figure 6b), implying a change in $\left(dT_g/dP\right)_{ex-situ} < 0$. These faster dynamics can be further accelerated by increasing the pressure during the densification process, as shown by data from a second HPAG produced under the same protocol but at 10 GPa (see main text, Figure 6a and 6b), which also exhibits lower thermal stability. As shown in Figure S9, the HPAG at 10 GPa exhibits more pronounced high-pressure effects on the FDSC, with a smaller endothermic peak above $T_g$ compared to the HPAG at 5 GPa.

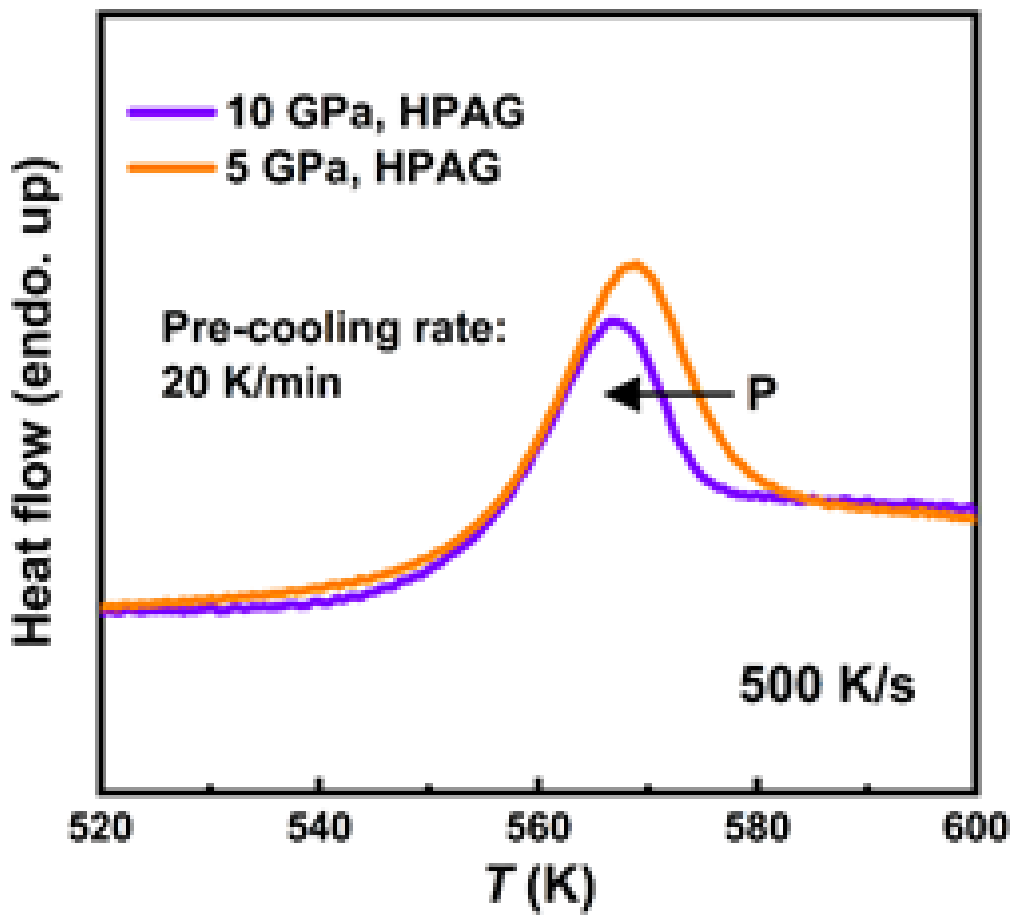

**Figure S9. Increased rejuvenation under pressure.** HPAGs compressed at 0.98$T_g$(5 GPa) and 5 and 10 GPa show enhanced rejuvenation at higher compression pressure.